# এনট্যাঙ্গলমেন্ট কী এবং কেন

উজ্জ্বল সেন

হরীশ-চন্দ্র রিসর্চ ইনস্টটুট, হোমী ভাভা ন্যাশনল ইনস্টটুটের অন্তর্গত,
ছত্নাগ রোড, ঝুঁসি, এলাহাবাদ ২১১ ০১৯, ভারতবর্ষ (ইন্ডিয়া)

What and why of entanglement

Ujjwal Sen
Harish-Chandra Research Institute, A CI of Homi Bhabha National Institute, Chhatnag Road, Jhunsi, Allahabad 211 019, India

**Abstract:** We present here a brief discussion, in Bangla (Bengali), on what is entanglement and why it is interesting.

### সংক্ষেপে

এখানে আমরা এনট্যাঙ্গলমেন্ট কী তা জানতে চেষ্টা করব। এনট্যাঙ্গলমেন্টের সংজ্ঞা তো জানবই, কিন্তু তার সঙ্গে এও জানার চেষ্টা করব যে কোয়ান্টম স্টেটের এই বিশেষ গুণ নিয়ে মাতামাতির দরকারটা কী।

## ১। গোড়ার কথা

আইনস্টাইন-পোডোলস্কি-রোজেনের বিখ্যাত সেই পেপর [1] এবং অন্যান্য কিছু রিসর্চ পেপর থেকে, ১৯৩৫ সাল নাগাদ, বোঝা যায় যে এনট্যাঙ্গলমেন্ট নামের একটা বৈশিষ্ট্য বা গুণ অনেক কোয়ান্টম স্টেটেই আছে। ভারতচন্দ্র যেমন দ্ব্যর্থবোধকভাবে লিখেছিলেন "কোনো গুণ নাহি তাঁর", অনেকটা সেরকমই এনট্যাঙ্গলমেন্ট নামক কোয়ান্টম স্টেটের গুণটা গুণ না দোষ, তা প্রথম চোটে তেমন বোঝা যায় নি। বেশ কিছু দশক ধরে এনট্যাঙ্গলমেন্টকে তাত্ত্বিকভাবে ব্যবহার করে বিভিন্ন প্যারাডক্স খোঁজারই রেওয়াজ ছিল। এনট্যাঙ্গলমেন্ট কোয়ান্টম স্টেটের একটা এমন বৈশিষ্ট্য, যা ক্লাসিকল জগতে মোটেই পাওয়া যায় না। সুতরাং স্বাভাবিকভাবেই এনট্যাঙ্গল্ড স্টেট ব্যবহার করে এমন সব তাত্ত্বিক ঘটনার বর্ণনা করা হচ্ছিল, যেগুলোকে ক্লাসিকল জগতের নিরিখে "ভুতুড়ে" মনে হচ্ছিল।

১৯৯০ নাগাদ, কিছু গবেষক এই দৃষ্টিকোণটাকে আমূল পাল্টে দিলেন। তাঁরা যুক্তি দিলেন, যেহেতু এনট্যাঙ্গলমেন্ট কোয়ান্টম স্টেটের এমন একটা বৈশিষ্ট্য, যা ক্লাসিকল জগতে অমিল, সেহেতু সেটা ব্যবহার করে নির্ঘাত এমন কাজ করা যাবে, যা ক্লাসিকল জগতে অসাধ্য বা দুরূহ। আর সত্যিই এরকম বেশ কিছু কাজের খোঁজ পাওয়া যেতে থাকল। মজার ব্যাপার হল, আগে যেসব ঘটনাগুলোকে প্যারাডক্স মনে হয়েছিল, সেগুলোকে অনেক সময়েই নতুন দৃষ্টিকোণ থেকে দেখে, কোনো কাজে লাগানো যেতে থাকল।

এখানে ভুললে চলবে না যে অনেকটা এই একই সময়ে, একক কোয়ান্টম সিস্টেম গবেষণাগারে সম্ভাব্য হতে শুরু করেছিল, এবং সেটা এই দৃষ্টিকোণ বদলে সাহায্য করে থাকতে পারে। এনট্যাঙ্গলমেন্টকে যে ব্যবহার করা যেতে পারে, সেটা কেন ১৯৩৫-এর পরিবর্তে ১৯৯০-এ বোঝা গেল, তার উত্তর হয়ত এটাই [২]।

সে যাই হোক, এই ছোট রিভিউতে আমরা এনট্যাঙ্গলমেন্টের সংজ্ঞা তো জানবই, কিন্তু তার সঙ্গে জানব এর কিছু কেজো দিক।

## ২। আমরা যেভাবে এগোব

৩। পিওর ও মিক্সড কোয়ান্টম স্টেট
    মিশ্রণের পরিমাণ
    কনভেক্সিটি
    বৈধ কোয়ান্টম স্টেটের সেটটা ক্লোজ্‌ড
৪। দুই বা ততোধিক অংশের বস্তুর কোয়ান্টম বিবরণ
৫। বস্তুর কিছুটা অংশ যদি বেপাত্তা হয়ে যায়: আংশিক ট্রেস
৬। একাধিক দূরবর্তী পর্যবেক্ষক ও LOCC
    স্থানীয় যোগবিয়োগ
    স্থানীয় পরিমাপ
    ফোনালাপ
    স্থানীয় ইউনিটেরি অপরেশন
    য দফা চাই, ত দফা চালাও
৭। সেপারেবল স্টেট
    কিছু উদাহরণ
    সেপারেবল স্টেটের সেটটা কনভেক্স
    সেপারেবল স্টেটের সেটটা ক্লোজ্‌ড
৮। যারা সেপারেবল নয়, তারা এনট্যাঙ্গল্‌ড
৯। যখন অংশদের সম্পর্কে কিছুই জানি না, অথচ গোটা বস্তুর ব্যাপারে সব জানি: সর্বোচ্চ এনট্যাঙ্গলমেন্ট
১০। কোয়ান্টম টেলিপোর্টেশন
    কোয়ান্টম টেলিপোর্টেশন বনাম বিশেষ আপেক্ষিকতা
    কাঁচামালের ফর্দ
    কোয়ান্টম ইনফর্মেশন ক্লোন করা যায় না ও তার সাথে কোয়ান্টম টেলিপোর্টেশনের সম্পর্ক

**১১। থেকে ১৮। পরের ভার্শনে থাকবে।**

১১। পিওর এনট্যাঙ্গল্‌ড স্টেট
১২। পজিটিভ আংশিক ট্রান্সপোজ় (PPT) নির্ণায়ক
১৩। বাউন্ড এনট্যাঙ্গলমেন্ট

## ৩। পিওর ও মিক্সড কোয়ান্টম স্টেট

কোয়ান্টম তত্ত্ব দাবি করে যে প্রত্যেক বস্তুর জন্যে তাত্ত্বিকভাবে আমরা একটা হিলবার্ট স্পেস বরাদ্দ করতে পারি। যেকোন হিলবার্ট স্পেস সংজ্ঞায়িত করার জন্যে লেজুড় হিসেবে একটা ফীল্ড লাগে, এবং কোয়ান্টম তত্ত্ব দাবি করে, যে কোনো বস্তুর জন্যে নির্দিষ্ট করা হিলবার্ট স্পেসকে অনুষঙ্গ হিসেবে জটিল সংখ্যার ফীল্ডটাকেই নিতে হবে। প্রত্যেক বস্তুর জন্যে আমাদের তার "ডিগ্রী অফ ফ্রীডমের" সংখ্যা জানতে হবে। অভিজ্ঞতা থেকে আমরা জানি যে এই সংখ্যা দুই বা তার চেয়ে বেশি কোনো পূর্ণসংখ্যা হয়, বা $\aleph_0$ হয়। কোয়ান্টম মেক্যানিক্সের চাহিদা, যে এটাই ওই বস্তুর হিলবার্ট স্পেসের ডামেনশন হতে হবে।

যে হিলবার্ট স্পেসের ডামেনশন, হয় (ফাইনাইট) পূর্ণসংখ্যা ($\geq 2$), নতুবা $\aleph_0$, তাকে সেপারেবল হিলবার্ট স্পেস বলে। মনে রাখতে হবে যে এর সাথে নিচে বর্ণিত সেপারেবল স্টেটের কোনো সম্পর্ক নেই।

এখানে মনে রাখা ভাল যে বস্তু কী, সে ব্যাপারে কোয়ান্টম তত্ত্ব টু শব্দ করে না। ইলেক্ট্রোম্যাগনেটিক ফীল্ডের একটা মোড, একটা বিড়াল, একটা রুমাল, সবই বস্তু হতে পারে। একইভাবে কোনো বস্তুর ডিগ্রী অফ ফ্রীডম কত, সেটাও সেই বস্তু সম্পর্কে আমাদের অভিজ্ঞতা থেকেই জানতে হবে।

কোয়ান্টম তত্ত্ব দাবি করে যে কোনো বস্তুর প্রত্যেক অবস্থার জন্যে একটা করে কোয়ান্টম স্টেট থাকবে। এই কোয়ান্টম স্টেট হবে সেই বস্তুর জন্যে বরাদ্দ হিলবার্ট স্পেসের ওপর সংজ্ঞায়িত করা একটা (হার্মিশিয়ান) পজিটিভ অপরেটর যার ট্রেস $= 1$. যদি এই অপরেটরের ম্যাট্রিক্স র‍্যাঙ্ক $= 1$ হয়, তাহলে আমরা বলি যে ওই কোয়ান্টম স্টেটটা "পিওর"। তা নাহলে ওই স্টেটটা "মিক্সড"। বস্তুর স্টেট পিওর হলে, তাকে আমরা তার হিলবার্ট স্পেসের একটা ভেক্টরের প্রজেক্টরের আকারে লিখতে পারি। আর তখন, আমরা বস্তুটার ওই অবস্থাকে প্রতিনিধিত্ব করার জন্যে প্রজেক্টরটা বা ভেক্টরটা - যেকোন একটা - নিতে পারি। ভেক্টরটা নিলে মনে রাখতে হবে যে একটা (পদার্থবিজ্ঞান অনুসারে) অনাবশ্যক এবং অর্থহীন ফেজ জুড়ে

গেছে বস্তুর অবস্থার বর্ণনায়। মিক্সড স্টেটের ম্যাট্রিক্স র্যাঙ্ক অবশ্যই যে হিলবার্ট স্পেসের ওপর সে সংজ্ঞায়িত, তার ডামেনশন অব্দি হতে পারে।

**মিশ্রণের পরিমাণ:** কোনো একটা মিক্সড স্টেটে মিশ্রণের পরিমাণ কতটা, সেটা বিভিন্ন ভাবে ভাবা যায়। তাদের মধ্যে একটা হলো ফন নয়মান এনট্রপি। কোনো একটা কোয়ান্টম স্টেট $\rho$-এর ফন নয়মান এনট্রপি হলো [3]

$$S(\rho) = -\text{tr}\rho \log_2 \rho.$$

এখানে লগারিদমের বেস হিসেবে আমরা 2 নিয়েছি, যেটা একরকম যদৃচ্ছ মনোনয়ন। কোয়ান্টম ইনফর্মেশনের বাজারে এই চয়নই চলে বেশি। দেখাই যাচ্ছে যে যেকোন পিওর স্টেটের ফন নয়মান এনট্রপি শূন্য হবে, যেটা কিনা এই বৈশিষ্ট্যের সর্বনিম্ন মান। সবচেয়ে বেশি মান হবে সম্পূর্ণরূপে ডিপোলারাইজ্ড স্টেটের। এখানে আমরা $\lim_{x\to 0+} x \log_2 x = 0$ ব্যবহার করেছি।

**কনভেক্সিটি:** দুটো কোয়ান্টম স্টেট মেশালে যা পাওয়া যায়, সেটা একটা বৈধ কোয়ান্টম স্টেট। খুবই উপকারী গুণ। ভাবা যাক, নাহলে কী হতো: কনভেক্সিটি সত্যি না হলে, একটা বস্তুকে কোনো দুটো স্টেটের একটাতে তৈরি করে ভুলে গেলে, বস্তুটা হঠাৎ উবে যেতে পারত।

কোনো একটা পিওর স্টেটকে আমরা কখনোই দুই বা তার চেয়ে বেশি পিওর স্টেটের মিশ্রণ থেকে পেতে পারি না। এটা প্রমাণ করতে আমরা যেকোন দুটো পিওর স্টেট নিয়ে, তাদের একটাকে $|0\rangle$ আর অন্যটাকে $|\psi\rangle = \alpha|0\rangle + \beta|1\rangle$ দিয়ে চিহ্নিত করতে পারি, যেখানে $|0\rangle$ আর $|1\rangle$ দুটো অর্থনর্মাল স্টেট, $\alpha, \beta$ দুটো জটিল রাশি, আর $|\alpha|^2 + |\beta|^2 = 1$. এদের দুটোকে যথাক্রমে $p$ আর $1-p$ প্রাবিলিটি দিয়ে মেশালে, আমরা পাব $\rho = p|0\rangle\langle 0| + (1-p)|\psi\rangle\langle\psi|$, যেটার আইগেনভ্যালু কষে বের করলে আমরা দেখতে পাব যে $\rho$-কে পিওর হতে গেলে, হয় $\beta = 0$ আর নাহয় $p = 0$ বা 1 হতে হবে।

দুটো পৃথক পিওর স্টেট মিশিয়েই যদি একটা পিওর স্টেট তৈরি না করা যায়, তাহলে তিনটে দিয়ে আরোই করা যাবে না, এরকমই মনে হয়। তবু প্রমাণ করে নেওয়া ভাল। এটা প্রমাণ করতে আমরা ফন নয়মান এনট্রপির কনকেভিটি ব্যবহার করতে পারি, যেটা বলে যে যেকোন দুটো কোয়ান্টম স্টেটকে মেশালে, মিশ্রনের ফন নয়মান এনট্রপি, গড় এনট্রপির চেয়ে বেশি হবে:

$$S(p\rho_1 + (1-p)\rho_2) \geq pS(\rho_1) + (1-p)S(\rho_2),$$

যেখানে $\rho_1$ আর $\rho_2$ হলো দুটো কোয়ান্টম স্টেট আর তাদের যথাক্রমে $p$ আর $1-p$ প্রাবিলিটি দিয়ে মেশানো হয়েছে।

এখন $\rho_1, \rho_2, \rho_3$ যদি তিনটে আলাদা আলাদা পিওর স্টেট হয়, এবং তাদের যথাক্রমে $p_1, p_2, p_3$ প্রাবিলিটি দিয়ে মেশানো হয়, তাহলে আমরা গোটা মিশ্রণটাকে প্রথম দুটোর মিশ্রণের সঙ্গে তৃতীয়টা মিশিয়েছি বলে ভাবতে পারি। আমরা ইতিমধ্যেই দেখেছি যে দুটো আলাদা পিওর স্টেট মেশালে একটা মিক্সড স্টেট তৈরি হয়। এবার ফন নয়মান এনট্রপির কনকেভিটি ব্যবহার করলেই আমরা দেখতে পাব যে তিনটে কোয়ান্টম স্টেট (কোনো প্রাবিলিটি শূন্য না করে) মেশালে আমরা কিছুতেই কোনো পিওর কোয়ান্টম স্টেট পেতে পারি না।

একইভাবে আমরা দেখতে পারব যে চার বা তার চেয়ে বেশি কোয়ান্টম স্টেট মেশালেও পিওর কোয়ান্টম স্টেট পেতে পারি না।

সুতরাং, দুটো বা ততোধিক ভিন্ন পিওর স্টেট মেশালে অবশ্যই একটা মিক্সড স্টেট পাওয়া যাবে। গোটাকয়েক মিক্সড স্টেট মেশালেও মিক্সড স্টেটই পাওয়া যাবে। কিন্তু আমরা নিচে দেখব যে একটা মিক্সড স্টেটকে আমরা একটা দু-অংশের (বা তার বেশি অংশের) বস্তুর পিওর স্টেট থেকেও পেতে পারি। অর্থাৎ, মিক্সড স্টেট পাওয়ার জন্যে কিছু পিওর স্টেট নিয়ে, তাদের মিশ্রণ করাটা আবশ্যক নয়। আর তাই, পিওর আর মিক্সড স্টেটদের বাংলায় অবিমিশ্র

আর মিশ্র স্টেট বা খাঁটি আর ভেজাল স্টেট বলার তেমন কারণ নেই। ইংরেজিতেও ওগুলোকে পিওর এবং মিক্সড স্টেট বলার দরকার একই কারণে নেই, কিন্তু সেটা বদলানো হয়ত একটু শক্ত।

**বৈধ কোয়ান্টম স্টেটের সেটটা ক্লোজ্ড:** কোয়ান্টম স্টেটের সেটটার বাউন্ডারি ভ্যালিড কোয়ান্টম স্টেট দিয়েই তৈরি। বাউন্ডারিতে আছে সেই সমস্ত স্টেট যাদের ম্যাট্রিক্স র‍্যাঙ্ক, স্পেসের ডামেনশনের থেকে এক বা আরো বেশি কম। সমস্ত পিওর স্টেট অবশ্যই আছে বাউন্ডারিতে।

## ৪। দুই বা ততোধিক অংশের বস্তুর কোয়ান্টম বিবরণ

বিভিন্ন বস্তুর সমষ্টি অবশ্যই আরেকটা বস্তু। এই সমষ্টির প্রত্যেক অংশের জন্যেই একটা করে হিলবার্ট স্পেস আছে। তিনটে বস্তু নিলে, এবং তাদের হিলবার্ট স্পেসগুলোকে যথাক্রমে $H_1$, $H_2$, $H_3$ নাম দিলে, ওই তিন বস্তুর সমষ্টির হিলবার্ট স্পেস হবে $H_1 \otimes H_2 \otimes H_3$. শেষোক্ত এই টেনসর-প্রডাক্ট হিলবার্ট স্পেস অবশ্যই একটা পুরদস্তুর হিলবার্ট স্পেস।

আর সেই হিলবার্ট স্পেসে আবার পিওর এবং মিক্সড স্টেট, ও তাদের কনভেক্সিটি ও ক্লোজ্ডনেস, একইভাবে সংজ্ঞায়িত ও ব্যবহার করা যাবে।

যদিও বললুম তিনের জন্যে, অবশ্যই ওই একই ব্যাপার দুই, চার, তেরো, সতেরো, সবার জন্যেই সত্যি।

## ৫। বস্তুর কিছুটা অংশ যদি বেপাত্তা হয়ে যায়: আংশিক ট্রেস

তিনজন খেলছিল আর তাদের প্রত্যেকের কাছে একটা করে জিনিস ছিল। ওদের নাম অরুণা, বরুণা, আর কিরণ। চলছিল ভালই, কিন্তু হঠাৎ কিরণ বলল আর খেলবে না। শুধু তাই নয়, সেই থেকে কিরণের সাথে আর যোগাযোগও করা যাচ্ছে না। এবার তাহলে অরুণা আর বরুণার কাছে থাকা জিনিসদুটোর স্টেটটা কি আমরা অরুণা-বরুণা-কিরণের স্টেটটা থেকে বের করতে পারব? এখন, কোয়ান্টম তত্ত্বের স্টেট বা অবজ়ারভবল জানতে হলে আমাদের হাতিয়ার কেবলমাত্র কোয়ান্টম পরিমাপ। সুতরাং, অরুণা-বরুণার স্টেটটা এমন হতে হবে, যাতে তাদের অবজ়ারভবলের ফলাফল কিরণ কিছুই না করলে এবং তিনজনেই খেললে যা হতো, এখনও তাই হয়। আংশিক ট্রেসের মাধ্যমে এটাই সাধিত হয়। অর্থাৎ, আমরা দেখাতে পারি যে $\rho_{ABK}$ যদি অরুণা-বরুণা-কিরণের কাছে থাকা একটা ত্রিপাক্ষিক কোয়ান্টম স্টেট হয় আর $O_{AB}$ যদি অরুণা-বরুণার অংশগুলোর ওপর একটা অবজ়ারভবল (হার্মিশিয়ান আপরেটর) হয়, তাহলে

$$\text{tr}_{ABK}\left(O_{AB} \otimes \mathbb{I}_K \rho_{ABK}\right) = \text{tr}_{AB}\left(O_{AB}\rho_{AB}\right),$$

হবে, যেখানে

$$\rho_{AB} = \text{tr}_K\left(\rho_{ABK}\right). \tag{1}$$

অতএব, $\{|k\rangle\}_k$ যদি কিরণের কাছে থাকা বস্তুর হিলবার্ট স্পেসের একটা সম্পূর্ণ এবং অর্থনর্মাল বেসিস হয়, তাহলে

$$\rho_{AB} = \sum_k \langle k|\rho_{ABK}|k\rangle. \tag{2}$$

সমীকরণ (1) বা (2)-এর অপরেশনকে আমরা আংশিক ট্রেস বলব। এখানে এবং অন্যত্র আমরা $\mathbb{I}_X$ বলতে $X$ নামক বস্তুর হিলবার্ট স্পেসের ওপর আইডেন্টিটি অপরেটর বুঝব।

এই একইভাবে আমরা একটা পাঁচ-অংশের জিনিসের থেকে দুটো বা চার-অংশের বস্তু থেকে তিনটে অংশ বাদ দিতে পারি।

আংশিক ট্রেস করে দেওয়ার পর, যে স্টেটটা পেলুম, সেটাকে আমরা আংশিক কোয়ান্টম স্টেট বলব। যেমন, $\rho_{AB}$ হলো $\rho_{ABK}$-র $AB$-অংশের আংশিক কোয়ান্টম স্টেট।

## ৬। একাধিক দূরবর্তী পর্যবেক্ষক ও LOCC

একটা দু-অংশের বস্তুর কথা ভাবা যাক। এই বস্তুটার একটা অংশ আছে আকবরের কাছে, অন্যটা আছে বীরবলের কাছে। আকবর আর বীরবলকে আমরা অনেকসময় $A$ আর $B$ বলেও সম্বোধন করব। ওদের দুজনের অবস্থান এক অন্যের থেকে দূরে হতে পারে। ওদের হেফাজতে থাকা বস্তুটার অংশগুলো ওরা ওদের আশপাশ থেকে জোগাড় করে থাকতে পারে। আবার এমনও হতে পারে যে ওই অংশগুলো একসাথে অন্য কারোর কাছে ছিল, এবং সে ও-দুটোর ওপর তখন কোনো ইন্টারাকশন হ্যামিলটোনিয়ান প্রয়োগ করেছিল, এবং তারপর আকবর আর বীরবলদের একটা করে অংশ পাঠিয়ে দিয়েছে।

যেভাবেই ওরা - মানে আকবর আর বীরবল - জিনিসদুটো পেয়ে থাকুক, এবার আমরা বুঝতে চেষ্টা করব, ওই অংশদুটোর ওপর ওরা কী কী করতে সক্ষম।

**স্থানীয় যোগবিয়োগ:** ধরা যাক যে আকবর-বীরবলের কাছে যে স্টেটটা আছে, সেটা $\rho_{AB}$, আর ওদের হিলবার্ট স্পেসগুলো হলো যথাক্রমে $H_A$ আর $H_B$। আকবর তার ল্যাবের আশ-পাশ থেকে একটা বাড়তি জিনিস নিয়ে, সেটাকে এই স্টেটের লেজুড় করে নিল। একইভাবে বীরবল একটা বাড়তি জিনিস তার গবেষণাগারের চারপাশ থেকে জোগাড় করতে পারে। আবার তারা দুজনেই তাদের নিজের নিজের অংশগুলোর কিছু কিছু টুকরো ফেলে দিতে পারে। তো ধরা যাক যে আকবর তার ল্যাবের কাছাকাছি থেকে একটা জিনিস নিল, যেটার হিলবার্ট স্পেস হলো $H_A^+$, আর যেটার স্টেট হলো $\rho_A^+$। আর বীরবল, $\rho_{AB}$-র ভেতর তার অংশ থেকে কিছুটা ফেলে দিল, যেটার হিলবার্ট স্পেস $H_B^-$, আর ওর কাছে যে টুকরোটা পড়ে রইল, সেটার হিলবার্ট স্পেস হলো $H_B{}'$। সুতরাং, $H_B = H_B{}' \otimes H_B^-$। আর আকবরের কাছে এখন যা আছে, তার হিলবার্ট স্পেস হলো $H_A \otimes H_A^+$, যার নাম দেয়া যাক $H_A{}'$। তাহলে, এখন ওদের দুজনের কাছে যে বস্তুটা আছে, তার হিলবার্ট স্পেস হলো $H_A{}' \otimes H_B{}'$, আর তার স্টেট হলো

$$\rho_{A'B'}^{(1)} = \text{tr}_{H_B^-}(\rho_{AB}) \otimes \rho_A^+.$$

এরপর, আকবর ও বীরবল ওদের নিজেদের অংশগুলোর ওপর কোয়ান্টম তত্ত্বে অনুমোদিত যেকোন কার্যকলাপ চালাতে পারে।

**স্থানীয় পরিমাপ:** এই ফেলে দেয়া আর জুড়ে নেয়া ছাড়া, কোয়ান্টম তত্ত্ব মাত্র দুটো কাজ অনুমোদন করে। ইউনিটেরী অপরেশন আর কোয়ান্টম পরিমাপ। ধরা যাক যে আকবর তার অংশের ওপর একটা কোয়ান্টম পরিমাপ করল। এই অংশের হিলবার্ট স্পেস হলো $H_A{}'$। সুতরাং, আকবর তার অংশের ওপর একটা অর্থাগনল প্রোজেকশন অপরেটরের সম্পূর্ণ সেটের পরিমাপ করতে পারবে। ধরা যাক যে তেমন একটা সেট হলো $\{P_i\}_{i=1}^n$। এই সে-টের উপাদানগুলো প্রজেক্টর, অর্থাৎ $P_i^2 = P_i \;\forall i$. এই প্রজেক্টরগুলো অর্থাগনল, অর্থাৎ $P_i P_j = \delta_{ij} \;\forall i \neq j$, আর তাই $n \leq \dim H_A{}'$। আর সেটটা সম্পূর্ণ, অর্থাৎ $\sum_{i=1}^n P_i = \mathbb{I}_{H_A{}'}$. যদি প্রত্যেকটা প্রজেক্টরের র‍্যাঙ্ক $= 1$ হয়, তাহলে $n = \dim H_A{}'$ হবে। ধরা যাক আকবর

দেখল যে তার পরিমাপযন্ত্র $i = 5$ দেখাচ্ছে। এটা ঘটার প্রাবিলিটি, বর্ন নিয়ম অনুসারে,

$$p_5 = \text{tr}\left(P_5 \otimes \mathbb{I}_{H_{B'}} \rho^{(1)}_{A'B'}\right),$$

আর তখন আকবর-বীরবলের কাছে যে স্টেটটা পড়ে থাকবে, সেটা হলো

$$\rho^{(2)}_{A'B'} := \frac{1}{p_5} P_5 \otimes \mathbb{I}_{H_{B'}} \rho^{(1)}_{A'B'} P_5 \otimes \mathbb{I}_{H_{B'}}.$$

এরপর আকবর $\rho^{(2)}_{A'B'}$-তে তার অংশের থেকে কিছুটা ফেলে দিতে পারে, বা কোনো অতিরিক্ত টুকরো জুড়েও নিতে পারে। সেসব করার পর আকবর যা পাবে, আমরা তার আর কোনো নতুন নাম দিচ্ছি না - ওই $\rho^{(2)}_{A'B'}$ বলেই তাকে ডাকছি।

**ফোনালাপ:** এইবার আমরা একটা নতুন ব্যাপার মেনে নেব। আমরা ধরে নেব যে আকবর আর বীরবল তাদের নিজেদের ভেতর ফোনে (বা অন্য কোনো ভাবে, যেমন চেঁচিয়ে বা চিঠি লিখে) কথাবার্তা বলতে পারে, যার মারফত তারা একে অন্যকে বলতে পারে তাদের পরিমাপের ফল কী হয়েছে। আমরা এও ধরে নেব যে এই ফল জানার পর, যাকে সেটা বলা হলো, সে এই ফলাফলের ওপর ভিত্তি করে তার ভবিষ্যৎ কর্মসূচি ঠিক করতে পারবে, অর্থাৎ এর পর সে কোন ইউনিটেরী আপরেটর প্রয়োগ করবে বা কোন প্রজেক্টরের সেট ব্যবহার করে পরিমাপ করবে, সেটা ঠিক করতে পারবে। সত্যি সত্যি এইরকম করতে গেলে কালঘাম ছুটে যেতে পারে, কিন্তু আমরা সেসব দিকে মন দেব না। ওপরে আমরা যে পরিমাপের কথা বলেছি, তাতে আকবর $i = 5$ পাওয়ার পর, আকবর জানল যে গোটা স্টেটটা রয়েছে $\rho^{(2)}_{A'B'}$-এ আর তার স্টেটটা রয়েছে $\text{tr}_{B'}\left(\rho^{(2)}_{A'B'}\right)$-এ। এই $i = 5$ পাওয়ার খবর আকবর যদি বীরবলকে ফোন করে বলে দেয়, তাহলে বীরবল জানবে যে গোটা স্টেটটা আছে ওই $\rho^{(2)}_{A'B'}$-এ আর তার স্টেটটা আছে $\text{tr}_{A'}\left(\rho^{(2)}_{A'B'}\right)$-এ। যদি আকবর কোনো খবর না পাঠায়, বা পাঠাতে অপারগ হয়, তাহলে বীরবলের কাছে ব্যাপার অন্যরকম হবে। যদি বীরবল নিশ্চিত হতে পারে যে পরিমাপটা করা হয়েছে, তাহলে ও জানবে যে গোটা স্টেটটা আছে $\rho^{(2)}_{A'B'}$-এ, যা কিনা পরিমাপের আগের স্টেট, অর্থাৎ $\rho^{(1)}_{A'B'}$-এর থেকে আলাদা। কিন্তু এই আলাদা হওয়াটা বীরবল কোনো পরিমাপ করে জানতে পারবে না, কারণ

$$\text{tr}_{A'}\left(\sum_{i=1}^{n} P_i \otimes \mathbb{I}_{H_{B'}} \rho^{(1)}_{A'B'} P_i \otimes \mathbb{I}_{H_{B'}}\right) = \text{tr}_{A'}\left(\rho^{(1)}_{A'B'}\right),$$

অর্থাৎ বীরবলের দিকে, পরিমাপের পরের গড় স্টেট আর পরিমাপের আগের স্টেট হুবহু এক।

**স্থানীয় ইউনিটেরি অপরেশন:** তো ধরা যাক যে বীরবল আকবরের কাছে খবর পেল যে পরিমাপের ফল $i = 5$ হয়েছে, এবং কোন প্রজেক্টরের সেট ব্যবহার করে পরিমাপ করা হয়েছে, সেটাও সে আগে থেকে জানে। এই খবর পাওয়ার পর, বীরবল কোনো নতুন টুকরো তার ল্যাবের আশপাশ থেকে জুড়ে নিতে পারে নিজের অংশে, বা নিজের অংশ থেকে কিছুটা ফেলে দিতেও পারে। আগের বারের মতোই, এটারও আমরা আর কোনো নতুন নাম না দিয়ে, আবার $\rho^{(2)}_{A'B'}$ বলেই ডাকব। এইবার বীরবল স্থির করল যে সে ওই দু-অংশের বস্তুটার, তার অংশের ওপর, একটা ইউনিটেরি আপরেটর $U_5$ প্রয়োগ করবে। এর ফলে গোটা স্টেটটা যে স্টেটে গিয়ে দাঁড়াবে, তা হলো

$$\rho^{(3)}_{A'B'} = \mathbb{I}_{H_{A'}} \otimes U_5 \; \rho^{(2)}_{A'B'} \; \mathbb{I}_{H_{A'}} \otimes U_5^{\dagger}.$$

যেহেতু $U_5$ একটা ইউনিটেরি আপরেটর, যেটা কিনা $H_{B'}$ হিলবার্ট স্পেসের ওপর সংজ্ঞায়িত, তাই $U_5^\dagger U_5 = \mathbb{I}_{H_{B'}}$ হবে।

**য দফা চাই, ত দফা চালাও:** যে দু-অংশের বস্তটা দুজনে পেয়েছিল, তার ওপর আকবর কিছু করা শুরু না করে বীরবল শুরু করতে পারে। আকবর শুরুতে পরিমাপ না করে ইউনিটেরি অপরেশন করতে পারে। আকবরের পরিমাপের পর বীরবল আবার পরিমাপ করতে পারে, এবং তার ফল এস-এম-এস করে আকবরকে জানাতে পারে। এবং এইরকম চলতেই থাকতে পারে। যতবার খুশি।

এই সবকিছু করাটাকে, সংক্ষেপে LOCC বলে। এখানে আমরা শুধু দু-অংশের বস্তু নিয়ে আলোচনা করেছি। একইভাবে, দুয়ের বেশি অংশের বস্তুর স্টেটের ওপরেও LOCC প্রয়োগ করা যায়। LOCC হলো local (quantum) operations and classical communication-এর আদ্যাক্ষর।

## ৭। সেপারেবল স্টেট

দুজন দূরে দূরে থাকা পর্যবেক্ষকের কথা ধরা যাক, যাদের আমরা আবার আকবর আর বীরবল বলে ডাকব, আর যারা শুধুমাত্র LOCC ব্যবহার করতে পারে বলে ধরে নেব। আকবর তার নিজের ল্যাবের আশপাশ থেকে একটা বস্তু জোগাড় করেছে। আমরা ধরে নিতে পারি যে এই বস্তুটা কোনো একটা পিওর স্টেটে আছে। যদি সেটা পিওর স্টেটে না থেকে থাকে, তাহলে আমরা ধরে নেব যে আকবর তার বস্তুটার স্থানীয় পরিমণ্ডল থেকে (অর্থাৎ তার গবেষণাগারের আশপাশ থেকে) কিছু টুকরোকে তার বস্তুর অংশ হিসেবে বিবেচনা করবে, এবং এই টুকরোর সংখ্যা ও পরিধি বাড়াতে থাকবে, যতক্ষণ না তার কাছে একটা এমন বস্তু থাকে, যার স্টেট পিওর হয়। এই পিওর স্টেটকে আমরা $|\psi_1\rangle$ নাম দেব, এবং সেটা যে হিলবার্ট স্পেসের উপাদান (অর্থাৎ ভেক্টর), তাকে আমরা $H_A$ বলব। একইভাবে আমরা বীরবলের জোগাড় করা বস্তুর স্টেটকে $|\phi_1\rangle$ বলব, এবং তার হিলবার্ট স্পেসকে $H_B$ বলব। দুজনের কাছে থাকা এই দুটো বস্তুকে আমরা একটা দু-অংশের একক বস্তু হিসেবে ভাবব। সুতরাং, এই দু-অংশের বস্তুটার স্টেট হলো $|\psi_1\rangle \otimes |\phi_1\rangle$, এবং সেটা $H_A \otimes H_B$-এর উপাদান।

এখানে আমরা ধরে নিয়েছি যে আকবর আর বীরবল শুধুমাত্র LOCC ব্যবহার করতে পারে। আর তাই তারা তাদের দু-অংশের বস্তুর ওপর কোনো ইন্টারাকশন হ্যামিলটোনিয়ান প্রয়োগ করতে পারবে না। এখানে বলে রাখা যাক যে ''এনট্যাঙ্গলমেন্ট স্যুয়্যাপিং'' বলে একটা মজাদার ব্যাপার আছে, যেখানে আকবরের সঙ্গে চারুর আর বীরবলের সাথে দীপিকার দেখা হয়, এবং তার আগে চারু-দীপিকার সাক্ষাৎ হয়, কিন্তু আকবর-বীরবলের কোনওদিন দেখা হয় না, আর এই সবের পর আকবর-বীরবলের কাছে এমন একটা স্টেট থেকে যায়, যেটা আকবর-বীরবল, চারু-দীপিকার সাহায্য ছাড়া, শুধু LOCC মারফত তৈরি করতে পারত না [4]। তা আমরা এখানে ধরে নেব যে আকবর-বীরবলের সাথে এরকম কোনো চারু-দীপিকার মোলাকাত হয় নি।

যেহেতু আমরা ধরে নিয়েছি যে আকবর আর বীরবল LOCC ব্যবহার করতে পারে, তাই তারা ফোনে কথা বলতেও পারবে। আর তাই, তারা যখন $|\psi_1\rangle \otimes |\phi_1\rangle$ তৈরি করবে, তারা সেটা নিজেদের ভেতর আলোচনা করে নিতে পারবে, এবং এমনও হতে পারে যে তারা মাঝে মাঝে $|\psi_1\rangle \otimes |\phi_1\rangle$ তৈরি করবে আর মাঝে মাঝে $|\psi_2\rangle \otimes |\phi_2\rangle$ তৈরি করবে, আর এগুলো তারা যথাক্রমে $p_1$ আর $p_2$ প্রাবাবিলিটি নিয়ে করবে। সুতরাং LOCC ব্যবহার করে আকবর আর বীরবল যে স্টেটগুলো তৈরি করতে পারবে, তাদের যেকোন একটাকে সবসময় এইভাবে

লেখা যেতে পারে:
$$\sigma_{AB} = \sum_i p_i |\psi_i\rangle\langle\psi_i| \otimes |\phi_i\rangle\langle\phi_i|. \tag{3}$$

এখানে $\{p_i\}_i$ একটা প্রাবাবিলিটি বিন্যাস। আর $|\psi_i\rangle \in H_A, |\phi_i\rangle \in H_B\ \forall i$.

ওপরের এই $\sigma_{AB}$ স্টেটগুলোকেই দু-অংশ বিশিষ্ট বস্তুদের সেপারেবল স্টেট বলে। সমীকরণ (3)-এর $\sum$ চিহ্নটা ডিস্ক্রীট বা কন্টিন্যুয়াস যোগফল ইঙ্গিত করতে পারে, আর তাই ওটা সাধারণ যোগফল না হয়ে ইন্টিগ্রেশন হতে পারে। সেপারেবল স্টেটকে এই আকারে যাঁরা প্রথম লেখেন তাদের মধ্যে অন্যতম হলেন ভেরানা [6]।

অবশ্যই এরকম হতে পারে যে দুয়ের বেশি পর্যবেক্ষক বিভিন্ন দূরে দূরে অবস্থিত গবেষনাগারে কাজ করছে, এবং শুধুমাত্র LOCC ব্যবহার করছে, তাদের মধ্যের অনেক-অংশের বস্তুটার ওপর। এক্ষেত্রে তারা যে স্টেটটা তৈরি করতে পারবে, সেটা এইরকম দেখতে হবে:
$$\sigma_{A_1 A_2 \ldots A_N} = \sum_i p_i |\psi_i^1\rangle\langle\psi_i^1| \otimes |\psi_i^2\rangle\langle\psi_i^2| \otimes \ldots \otimes |\psi_i^N\rangle\langle\psi_i^N|.$$

এখানে আমরা ধরে নিয়েছি যে $A_1, A_2, \ldots, A_N$ নামের $N$-জন পর্যবেক্ষক আছে যারা $N$টা দূরে দূরে অবস্থিত ল্যাবে আছে। তাদের কাছে একটা $N$-অংশের বস্তু আছে এবং তার ওপর ওরা শুধু LOCC প্রয়োগ করতে পারবে। আর পর্যবেক্ষক $A_j$-র কাছে থাকা অংশের হিলবার্ট স্পেস হলো $H_{A_j}, \forall j = 1, 2, \ldots, N$. সুতরাং $|\psi_i^j\rangle \in H_{A_j}\ \forall i$ আর $\forall j = 1, 2, \ldots, N$. আগের মতোই, $\{p_i\}_i$ একটা প্রাবাবিলিটি বিন্যাস।

দুয়ের বেশি পর্যবেক্ষকের ক্ষেত্রে গল্পটা অনেক বেশি জটিল, এবং তার কিছুটা আভাস আমরা শেষের দিকের একটা পরিচ্ছেদে পাব। কিন্তু এবার থেকে আমরা শুধুমাত্র দু-অংশের বস্তু নিয়েই কথাবার্তা বলব, যদি না স্পষ্টভাবে অন্যরকম বলা থাকে।

**কিছু উদাহরণ:** এবার কয়েকটা সেপারেবল স্টেটের উদাহরণ দেওয়া যাক, যারা প্রত্যেকেই কোনো না কোনো দু-অংশের বস্তুর স্টেট। $|00\rangle, |11\rangle, |0+\rangle, |++\rangle$, এরা সবাই সেপারেবল স্টেট। এদের যেকোন মিশ্রণও সেপারেবল। যথা,
$$\frac{1}{3}|00\rangle\langle 00| + \frac{2}{3}|11\rangle\langle 11|,$$
বা
$$\frac{1}{4}|11\rangle\langle 11| + \frac{3}{4}|0+\rangle\langle 0+|,$$
বা
$$\frac{1}{9}|00\rangle\langle 00| + \frac{8}{9}|++\rangle\langle ++|,$$
বা
$$p_1|00\rangle\langle 00| + p_2|11\rangle\langle 11| + p_3|0+\rangle\langle 0+| + p_4|++\rangle\langle ++|.$$

এখানে $|0\rangle$ আর $|1\rangle$ দুটো অর্থনর্মাল ভেক্টর। আর $|+\rangle = (|0\rangle + |1\rangle)/\sqrt{2}$. $\{p_1, p_2, p_3, p_4\}$ হলো একটা প্রাবাবিলিটি বিন্যাস।

**সেপারেবল স্টেটের সেটটা কনভেক্স:** এটা সহজেই দেখা যায় যে দুটো বা আরো বেশি সেপারেবল স্টেটের যেকোন মিশ্রণে আবার একটা সেপারেবল স্টেট গঠিত হবে। আমরা পরে দেখব যে সহজ এই ব্যাপারটা অতি গুরুত্বপূর্ণ একটা তথ্য। এখানে আমরা মনে রাখতে পারি যে সমস্ত কোয়ান্টম স্টেটের সেটটাও কনভেক্স।

***সেপারেবল স্টেটের সেটটা ক্লোজ্‌ড:*** এই ব্যাপারটার প্রমাণও বেশ সোজাসাপ্টা এবং বেশ গুরুত্বপূর্ণ। এবং আবার আমরা মনে রাখতে পারি যে সমস্ত কোয়ান্টম স্টেটের সেটটাও ক্লোজ্‌ড।

চিত্র 1-এ আমরা ওপরের বিষয়টার একটা স্থূল চিত্রণ দেওয়ার চেষ্টা করেছি।

## ৮। যারা সেপারেবল নয়, তারা এনট্যাঙ্গল্ড

ওপরে আমরা সেপারেবল স্টেটের উদাহরণ দেখেছি। এরকম উদাহরণের সহজেই সংখ্যাবৃদ্ধি করা যায়, আর এগুলোর গঠন থেকে সহজেই বোঝা যায় যে ওরা সেপারেবল। এসব করে কী হবে, সে প্রশ্নকে পাত্তা না দিয়েও আমরা অন্তত দুটো প্রশ্ন করতে পারি। সেপারেবল ব্যাতিত অন্য কোয়ান্টম স্টেট আছে কী? আর যদি থাকে, তাহলে কোনো একটা স্টেট সেপারেবল না সেপারেবল নয়, সেটা বুঝব কীভাবে?

প্রথম প্রশ্নের উত্তর "হ্যাঁ", আর দ্বিতীয়টার উত্তর "যদ্দূর জানি, এখনও কেউ জানে না"। প্রথম প্রশ্নোত্তর নিয়ে আরো কিছু কথা আমরা এখন বলব। দ্বিতীয় প্রশ্নটা এখন মুলতুবি রাখা হবে। অন্য পরিচ্ছেদে এ নিয়ে কিছু কথাবার্তা থাকবে।

এমন দু-অংশকে কোয়ান্টম স্টেট যে আছে, যারা সেপারেবল নয়, এটা দেখানোর জন্যে একটা উদাহরণই যথেষ্ট। আমরা এখানে দুটো উদাহরণ দেখবো, যাদের একটা পিওর আর অন্যটা মিক্সড।

প্রথমে দেখা যাক, পিওর সেপারেবল স্টেট কীরকম দেখতে হয়। আমরা আগেই দেখেছি যে কোনো পিওর স্টেটকেই কিছু স্টেটকে মিশিয়ে তৈরি করা যায় না। সুতরাং, কোনো সেপারেবল স্টেট যদি পিওর হয়, তাহলে সমীকরণ (3)-এর ডানদিকের যোগফলটাতে একটাই মাত্র টার্ম থাকবে। অর্থাৎ, যেকোন দু-অংশের সেপারেবল পিওর স্টেটকে আমরা সবসময় এইভাবে লিখতে পারব:

$$|\psi\rangle \otimes |\phi\rangle.$$

উল্লেখ্য যে পুরোমাত্রায় LOCC-র ব্যবহার, সেপারেবল স্টেট তৈরি করতে দরকার হয় না। সেপারেবল স্টেটের সাধারণ গঠনটা দেখলেই আমরা বুঝতে পারি যে তাদের তৈরি করতে শুধুমাত্র একমুখী LOCC-র ব্যবহারই যথেষ্ট। একমুখী LOCC বলতে আমরা বলতে চাইছি সেই সমস্ত LOCC-র কথা, যাদের নিষ্পন্ন করার জন্যে শুধু কোনো একজন পর্যবেক্ষকের কাছ থেকে অন্যজনের কাছে একবার এবং একমুখী ফোনালাপই যথেষ্ট। আবার, পিওর সেপারেবল স্টেট তৈরি করতে কোনো ফোনালাপের দরকারই নেই - শুধু LO-ই কাফি।

$|\psi\rangle \otimes |\phi\rangle$ ছাড়া অন্য পিওর স্টেট কি আছে একটা দু-অংশের টেনসর-প্রোডাক্ট হিলবার্ট স্পেসে? উত্তর হলো: আছে তো বটেই, একটা দু-অংশের টেনসর-প্রোডাক্ট হিলবার্ট স্পেসের প্রায় কোনো ভেক্টরই সেপারেবল নয়। প্রথমত, সিঙ্গলেট স্টেটটা নেওয়া যাক:

$$|\psi^-\rangle = \frac{1}{\sqrt{2}}(|01\rangle - |10\rangle).$$

এই স্টেটটা, যেটা $\mathbb{C}^2 \otimes \mathbb{C}^2$-র একটা ভেক্টর, যদি সেপারেবল হয়, তাহলে তাকে এইভাবে লেখা যাবে:

$$(a|0\rangle + b|1\rangle) \otimes (c|0\rangle + d|1\rangle).$$

একটু টানাহ্যাঁচড়া করলেই দেখা যাবে যে কোনো জটিল রাশির সমষ্টি $\{a, b, c, d\}$ পাওয়া যাবে না, যাতে সিঙ্গলেট স্টেটটা সেপারেবল স্টেটের আকারে লেখা যায়। সুতরাং, সিঙ্গলেট স্টেটটা সেপারেবল নয়। অর্থাৎ, একটা দু-অংশের বস্তু, যার দুটো অংশই দুই ডামেনশনের,

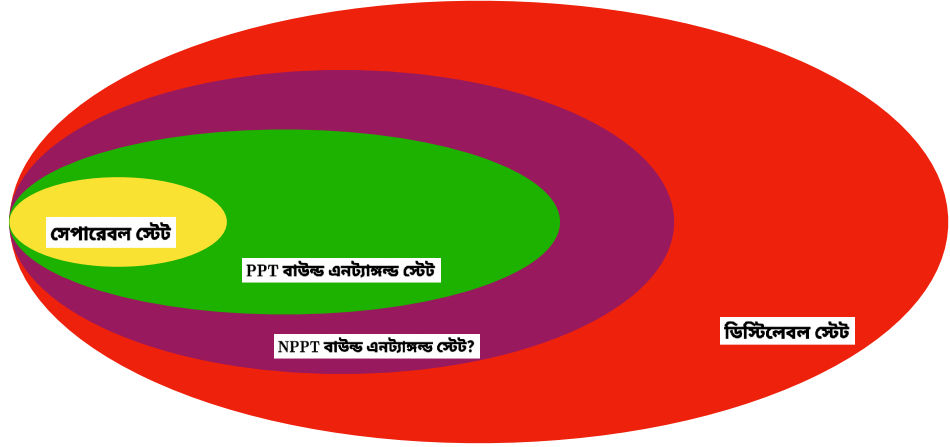

চিত্র 1: সবচেয়ে কম ডামেনশনের যে বস্তুতে এনট্যাঙ্গলমেন্ট দেখা যায়, সেটা হলো $\mathbb{C}^2 \otimes \mathbb{C}^2$. এখানকার স্টেটরা পনেরোটা বাস্তব সংখ্যার ওপর নির্ভরশীল। সুতরাং, এই ধরণের স্টেটদের কোনো সেটকে দুই- বা তিন-ডামেনশনের স্পেসে ফুটিয়ে তুলতে চেষ্টা করলে, প্রভূত অস্পষ্টতা এবং দ্ব্যর্থতা থেকে যাওয়া স্বাভাবিক। এটা মনে রেখেও আমরা সেপারেবল এবং এনট্যাঙ্গল্ড স্টেটদের একটা প্লেনার ছবিতে চিত্রিত করতে চেষ্টা করব। তো ধরা যাক আমাদের কাছে আছে একটা দু-অংশের বস্তু, যার একটা অংশ $d_A$ ডামেনশনের, আর অন্যটা $d_B$-র। এই বস্তুর সমস্ত কোয়ান্টম স্টেটদের আমরা একটা উপবৃত্তের ভেতরের বিন্দুগুলোকে দিয়ে প্রতিনিধিত্ব করাব, আর এটাই ওপরের ছবির সবচেয়ে বড় উপবৃত্তটা, অর্থাৎ লাল, মেরুন, সবুজ, হলুদ মিলিয়ে গোটা অঞ্চলটা। আর হলুদ অংশটা প্রতিনিধিত্ব করছে সেপারেবল স্টেটদের সেটটাকে। এই দুটো সেটই কনভেক্স এবং ক্লোজ্‌ড। হলুদ এবং সবুজ মিলিয়ে যে অঞ্চলটা, সেটাও একটা কনভেক্স এবং ক্লোজ্‌ড খণ্ড, এবং এটা প্রতিনিধিত্ব করছে সমস্ত PPT স্টেটদের। শুধু সবুজ অঞ্চলটাতে রয়েছে PPT বাউন্ড এনট্যাঙ্গল্‌ড স্টেটরা। হলুদ, সবুজ, ও মেরুন মিলিয়ে যে অঞ্চলটা, সেটা ছবিতে কনভেক্স হিসেবে চিত্রিত হলেও, ওটা প্রতিনিধিত্ব করছে সেই সমস্ত স্টেটদের যারা ডিস্টিলেবল নয়, এবং এই স্টেটদের সেটটা কনভেক্স কিনা জানা নেই। শুধু তাই নয়, মেরুন অঞ্চলটাতে আছে সমস্ত NPPT বাউন্ড এনট্যাঙ্গল্‌ড স্টেটরা, যাদের একটাও উদাহরণ আজ অব্দি পাওয়া যায় নি। যে সমস্ত স্টেট ডিস্টিলেবল নয়, তাদের সেটটা অন্তত সেইসব ক্ষেত্রে ক্লোজ্‌ড, যেখানে $d_A$ বা $d_B$ বা দুটোই $= 2$. লাল অংশটাতে রয়েছে সমস্ত ডিস্টিলেবল স্টেট। $d_A$ আর $d_B$ দুটোই 2 হলে, বা কোনো একটা 2 আর অন্যটা 3 হলে, মেরুন ও সবুজ অংশদুটো থাকে না, আর তাই এক্ষেত্রে PPT নির্ণায়ক দিয়েই আমরা সেপারেবিলিটি প্রশ্নের সমাধান করতে পারি। **[এই ছবিকে পুরোপুরি বুঝতে এই লেখাটার পরের ভার্শনটাও দেখতে হবে।]**

সে এমন পিওর স্টেটে থাকতে পারে, যাকে সেপারেবল ফর্ম, $|\psi\rangle \otimes |\phi\rangle$-তে, লেখা যায় না। আর তাই, এই দু-অংশের বস্তুর এমন স্টেট হয়, যাকে LOCC দিয়ে তৈরি করা যাবে না।

$\mathbb{C}^2 \otimes \mathbb{C}^2$-র যেকোন একটা ভেক্টরকে এইভাবে লেখা যায়:

$$\alpha|00\rangle + \beta|01\rangle + \gamma|10\rangle + \delta|11\rangle, \tag{4}$$

যেখানে $\alpha, \beta, \gamma, \delta$ হলো চারটে জটিল সংখ্যা, যারা নর্মালাইজ্ড: $|\alpha|^2 + |\beta|^2 + |\gamma|^2 + |\delta|^2 = 1$. এই স্টেটটাকে যদি সেপারেবল ফর্ম, অর্থাৎ এক্সপ্রেশন (4)-এর মতো করে লিখতে হয়, তাহলে (4)-র স্টেটটার দুটো অংশেরই আংশিক স্টেটগুলোর ফন নয়মান এনট্রপি শূন্য হবে। (যেহেতু গোটা স্টেটটা পিওর, তাই দুদিকের ফন নয়মান এনট্রপি একই হবে।) এই আংশিক স্টেটগুলো পেতে আমাদের আংশিক ট্রেস ব্যবহার করতে হবে। $\alpha, \beta, \gamma, \delta$-রা আছে একটা ছয় ডামেনশনের স্পেসে, যেটা কিনা $\mathbb{R}^6$-এর একটা উপসেট। আংশিক স্টেটগুলোর ফন নয়মান এনট্রপি শূন্য হতে হলে, $\mathbb{R}^6$-এর এই উপসেটের ভেতর একটা সারফেস পাব। এই সারফেসেই সমস্ত পিওর সেপারেবল স্টেট আছে, আর তাই, পিওর স্টেটদের ভেতর থেকে র‍্যান্ডম-ভাবে একটা স্টেট চয়ন করলে, সেপারেবল স্টেট পাওয়ার প্রাবিলিটি শূন্য। অর্থাৎ, "প্রায় সব" পিওর স্টেটই সেপারেবল নয়।। কোনো একটা ঘটনা "প্রায় সর্বদা" হচ্ছে বলা হলে, তার মানে হলো সেই ঘটনাটা না ঘটার প্রাবিলিটি শূন্য। যদিও ঘটনাটা ঘটতে পারে। যেমন, একটা ধুলোর কণার একটা টেবলের ওপরের রেখার ওপর পড়ার প্রাবিলিটি শূন্য, যদি কণাটার আয়তন আর রেখাটার প্রস্থ শূন্য ধরে নিই, কিন্তু কণাটা রেখাটার ওপর পড়ার ঘটনাটা ঘটতে পারে। এই অর্থেই আমরা ওপরে লিখেছিলুম যে "একটা দু-অংশের টেনসর-প্রোডাক্ট হিলবার্ট স্পেসের প্রায় কোনো ভেক্টরই সেপারেবল নয়"। আমরা এটা শুধুমাত্র $\mathbb{C}^2 \otimes \mathbb{C}^2$-তেই দেখলুম। কিন্তু এই একই প্রমাণ আমরা যেকোন দু-অংশের বস্তুর পিওর স্টেটদের ক্ষেত্রেই করতে পারি।

মিক্সড স্টেটদের মধ্যে সেপারেবল নয় এমন স্টেট আছে কিনা জানার জন্যে আমরা বিভিন্ন ভাবে এগোতে পারি। আগামী একটা পরিচ্ছেদে আমরা ''PPT নির্ণায়কের'' মাধ্যমে এটা দেখব। এখানে আমরা সেপারেবল নয় এমন মিক্সড স্টেট খুঁজব একটু অন্যভাবে।

যেকোন দু-অংশের সেপারেবল স্টেট $\sigma_{AB}$-র ক্ষেত্রে আমরা দেখাতে পারি যে

$$S(\sigma_A) \leq S(\sigma_{AB}), \quad S(\sigma_B) \leq S(\sigma_{AB}), \tag{5}$$

অর্থাৎ গোটা বস্তুটার মিশ্রতা, তার প্রত্যেক অংশের মিশ্রতার চেয়ে কম হবে। এখানে,

$$\sigma_A = \text{tr}_B \sigma_{AB},$$

এবং একইভাবে $\sigma_B$ সংজ্ঞায়িত। এই বিবৃতির প্রমাণ এই পেপারে আছে: [5]।

কাকে সিঙ্গলেট স্টেট বলে সেটা আমরা আগেই জেনেছি। এই স্টেটের সঙ্গে যদি সম্পূর্ণ-ভাবে ডিপোলারাইজ্ড স্টেটটা, অর্থাৎ $\frac{1}{4}\mathbb{I}_2 \otimes \mathbb{I}_2$, মেশানো হয়, তাহলে সেই স্টেটটাকে ড়্যেয়ানা স্টেট বলে [6]:

$$p|\psi^-\rangle\langle\psi^-| + \frac{1-p}{4}\mathbb{I}_2 \otimes \mathbb{I}_2. \tag{6}$$

এখানে $\mathbb{I}_2$ হলো একটা দুই ডামেনশনের হিলবার্ট স্পেসের ওপর সংজ্ঞায়িত করা আইডেন্টিটি আপরেটর। ড়্যেয়ানা স্টেটটাকে একটা বৈধ কোয়ান্টম স্টেট হতে গেলে $-1/3 \leq p \leq 1$ হতে হবে। এটা প্রমাণ করার জন্যে আমাদের ড়্যেয়ানা স্টেটটার আইগেনভ্যালু বের করে দেখতে হবে, $p$-এর কোন এলাকার জন্য আইগেনভ্যালুগুলো ধনাত্মক থাকে। তবে, $p$ ঋণাত্মক হয়ে গেলে, ড়্যেয়ানা স্টেটটাকে আর সিঙ্গলেট এবং সম্পূর্ণরূপে ডিপোলারাইজ্ড স্টেটের মিশ্রণ ভাবা যাবে না।

তা এই আইগেনভ্যালুগুলো নির্ণয় করা হয়ে গেলে, আমরা ড্ভয়ানা স্টেটটার ফন নয়মান এনট্রপিও বের করে ফেলতে পারব। আর, যেকোন $p$-এর জন্যে, ড্ভয়ানা স্টেটের দুটো আংশিক স্টেটই $\frac{1}{2}\mathbb{I}_2$, যাদের প্রত্যেকটার ফন নয়মান এনট্রপি 1. এখন যদি আমরা দেখাতে পারি যে $p$-এর কোনো একটা পরিসরে, ড্ভয়ানা স্টেটটার ফন নয়মান এনট্রপি 1-এর চেয়ে কম, তাহলে সেই স্টেটগুলো সেপারেবল হতে পারবে না। কারণ আমরা আগেই বলেছি ((5)-এর অসমীকরণদুটো দ্রষ্টব্য) যে কোনো দু-অংশের সেপারেবল স্টেটের ক্ষেত্রে গোটা স্টেটটার ফন নয়মান এনট্রপি তার যেকোন অংশের ফন নয়মান এনট্রপির চেয়ে কম হতে পারে না। একটু চেষ্টাচরিত্র করলেই দেখা যায় যে $p \gtrsim 0.7476$ হলে, ড্ভয়ানা স্টেটটার ফন নয়মান এনট্রপি 1-এর চেয়ে কম হয়, এবং তাই তারা কেউ সেপারেবল নয়।

তাহলে আমরা দেখতে পেলুম যে প্রায় কোনো পিওর দু-অংশের স্টেট সেপারেবল নয়। (পিওর দু-অংশের বস্তুদের স্টেটদের জন্যে, অসমীকরণ (5) ব্যবহার করে আমরা তাত্ত্বিকভাবে সহজেই তাদের এনট্যাঙ্গলমেন্ট সনাক্ত করতে পারি।) তাছাড়া অসীম সংখ্যায় উপাদান আছে এমন মিশ্রড স্টেটের একটা সেট পেয়েছি যার উপাদানগুলো কেউ সেপারেবল স্টেট নয়। যে সমস্ত দু-অংশের স্টেট সেপারেবল নয়, তাদের আমরা এনট্যাঙ্গল্ড বলি।

শুনেছি, প্রাচীন কোন গ্রন্থে বলা আছে যে ব্রহ্ম কী তা জানা যায় না, এবং তাই তা কী নয়, সেকথা বলার চেষ্টা করা হয়েছে সেখানে [7]। এখানে কিন্তু পুরোপুরি সেরকম নয়। এনট্যাঙ্গল্ড অবস্থা কী, তার মধ্যে কোনো কবিতাসুলভ ধোঁয়াশা নেই। শুধু সংজ্ঞাটা নেতিবাচক। উপরন্তু, এনট্যাঙ্গল্ড স্টেটের সোজাসুজি সংজ্ঞা দেওয়ার চেষ্টা চলছে, এবং কিছু সাফল্য এসেছে। কিন্তু আমরা এখানে সেসব নিয়ে আলোচনায় যাব না।

## ৯। যখন অংশদের সম্পর্কে কিছুই জানি না, অথচ গোটা বস্তুর ব্যাপারে সব জানি: সর্বোচ্চ এনট্যাঙ্গলমেন্ট

ইনফর্মেশনের অনেক মাপকাঠি আছে। তাদের একেকটার একেকরকম উদ্দেশ্য ও ব্যবহার। কিন্তু মাপকাঠি যাই হোক না কেন, যে বস্তুকে আমরা পিওর স্টেট দিয়ে বর্ণনা করছি, তার সম্পর্কে সবকিছু জানি, আর যে জিনিসকে আমরা একটা সম্পূর্ণ ডিপোলারাইজ্ড স্টেটের মারফত বর্ণনা করছি, তার ব্যাপারে আমরা কিছুই জানি না, এদুটো ব্যাপার নিয়ে তেমন কোনো মতভেদ নেই। কারণ হিসেবে বলা যায় যে কোনো এক বস্তু যদি পিওর স্টেটে থাকে, তাহলে তার হিলবার্ট স্পেসে অন্তত একটা সম্পূর্ণ এবং অর্থনর্মাল বেসিস থাকবে, যে বেসিসে কোয়ান্টম পরিমাপ করলে আমরা বর্ন প্রাবিবিলিটিগুলো $\{1, 0, 0, \ldots\}$-এইরকম পাব। এই প্রাবিবিলিটি বিন্যাসের শ্যানন এনট্রপি শূন্য। সম্পূর্ণ ডিপোলারাইজ্ড স্টেট, তার নিজের হিলবার্ট স্পেসের আইডেন্টিটি অপরেটরের সঙ্গে সমানুপাতিক। কোনো বস্তু এই স্টেটে থাকলে, তাকে যেকোন সম্পূর্ণ এবং অর্থনর্মাল বেসিসে পরিমাপ করলে, সেই পরিমাপের বর্ন প্রাবিবিলিটি বিন্যাস $\{1/d, 1/d, \ldots\}$-এইরকম হবে, যেখানে $d$ হলো ওই বস্তুর হিলবার্ট স্পেসের ডামেনশন। এই প্রাবিবিলিটি বিন্যাসের শ্যানন এনট্রপি সর্বাধিক ($= \log_2 d$). কোনো প্রাবিবিলিটি বিন্যাস $\{p_i\}$-এর শ্যানন এনট্রপি হলো $-\sum_i p_i \log_2 p_i$. ফন নয়মান এনট্রপির মতো এখানেও লগারিদমের বেস 2 নেওয়া হলো।

এবার ধরা যাক, আমরা একটা দু-অংশের বস্তু নিলুম, যার একেকটা অংশ একটা কোয়ান্টম স্পিন-$1/2$. এই স্পিনদুটো দুজন মানুষের জিম্মায় রয়েছে, যাদের নাম (আবার) আকবর আর বীরবল। এও ধরে নিলুম যে সেই বস্তুটার কোয়ান্টম স্টেট হল সিঙ্গলেট স্টেট। এখন এই গোটা বস্তুটা একটা পিওর স্টেটে রয়েছে, এবং তাই আমরা ভাবতে পারি যে এই গোটা বস্তুটা সম্পর্কে আমরা ''সব জানি''। অর্থাৎ, কোয়ান্টম তত্ত্ব মেনে চলে যতটা জানা যায়, তার সবটা জানি। এদিকে এই দু-অংশের বস্তুটার প্রত্যেকটা অংশ রয়েছে একটা সম্পূর্ণ ডিপোলারাইজ্ড স্টেটে, কারণ $\text{tr}_A |\psi^-\rangle\langle\psi^-| = (1/2)\mathbb{I}_2 = \text{tr}_B |\psi^-\rangle\langle\psi^-|$. আর তাই, এই

দুই-স্পিনের বস্তুর যেকোন অংশ সম্পর্কে আমাদের ইনফর্মেশন শূন্য।

এরকম ব্যাপারস্যাপার ক্লাসিকল জগতে কস্মিনকালে হয় না। ওই দুটো স্পিনের ক্লাসিকল পিওর স্টেট মাত্র চারটে। যথা,

$$|00\rangle, |01\rangle, |10\rangle, |11\rangle.$$

ধরা যাক যে এই $|0\rangle$ আর $|1\rangle$) বলতে আমরা $z$-ডিরেকশনের আপ আর ডাউন স্টেটগুলোকে বলতে চেয়েছি, আর তাই এই ডিরেকশনের আপ আর ডাউন স্টেটগুলোকে ক্লাসিকল বলে দাবি করেছি, যেকোন একটা স্পিন-$1/2$-এর ক্ষেত্রে। এই ডিরেকশনের বাছাই অবশ্যই কোন নিয়ম মেনে নয়, এবং আমরা বিন্দাস অন্য কোন ডিরেকশন মনোনীত করতে পারতুম। কিন্তু একবার কোন একটা ডিরেকশনের আপ এবং ডাউন স্টেটদের ক্লাসিকল বলে দাবি করে দিলে, সেই আলোচনায় আর সেটা পাল্টানো চলবে না। এখানে আমরা টু-ডামেনশনল কোয়ান্টম বস্তু নিয়েই কথা বলছি। ডামেনশন বেশি হলে, কোন একটা বেসিসকে আমাদের ক্লাসিকল বলে দাবি করতে হবে। দুটো স্পিন-$1/2$-এর বস্তুটা এই চারটে স্টেটের যেকোন একটাতে থাকলে, একেকটা স্পিনও পিওর স্টেটেই থাকবে। ওই দুটো স্পিনের যেকোন ক্লাসিকল স্টেট, যদি তারা মিক্সড স্টেটও থাকতে পারে বলে ধরে নিই, তাহলে তাদের আমরা এইভাবে লিখতে পারি:

$$p_1 |00\rangle \langle 00| + p_2 |01\rangle \langle 01| + p_3 |10\rangle \langle 10| + p_4 |11\rangle \langle 11|,$$

যেখানে $\{p_i\}_{i=1}^4$ একটা প্রাবিবিলিটি বিন্যাস। এই স্টেটটা পিওর হওয়ার একমাত্র উপায় যেকোন তিনটে $p_i$ শূন্য হয়ে যাওয়া। এবং তখন দুটো অংশ আবার পিওর স্টেটে থাকবে।

সিঙ্গলেট স্টেটের ক্ষেত্রে ক্লাসিকল মেক্যানিক্সের এই বেএখতিয়ার হয়ে পড়াটাকে সিঙ্গলেট স্টেটের ভেতর ($\mathbb{C}^2 \otimes \mathbb{C}^2$-এর স্টেটদের মধ্যে) সর্বোচ্চ এনট্যাঙ্গলমেন্ট থাকার নির্দেশক ভাবতে পারি। কিন্তু আমরা পরে এও দেখব যে এই স্টেট ব্যবহার করে অনেক সময়েই এমন কাজ করা যায়, যা সমস্ত $\mathbb{C}^2 \otimes \mathbb{C}^2$-এর কোয়ান্টম স্টেটদের ভেতর সব থেকে বেশি সুবিধে দেয়, ক্লাসিকল জগতে সেই একই কাজ করার সাপেক্ষে।

সিঙ্গলেট স্টেটের যেসব গুণের জন্যে তার সর্বোচ্চ এনট্যাঙ্গলমেন্ট আছে বলে দাবি করছি, সেসবই যেকোন $U_A \otimes U_B |\psi^-\rangle$-এরও আছে, যেখানে $U_A$ এবং $U_B$ হলো $\mathbb{C}^2$-র ওপর দুটো ইউনিটেরি অপরেটর। সুতরাং এরা সকলেই $\mathbb{C}^2 \otimes \mathbb{C}^2$-র সর্বোচ্চ এনট্যাঙ্গলমেন্ট-ওলা স্টেট।

এতক্ষণ আমরা একটা দু-অংশের বস্তুর সর্বোচ্চ এনট্যাঙ্গলমেন্টের কথা বলছিলুম, যার দুটো অংশেরই ডামেনশন $2$। দুটো $d$-ডামেনশনের জিনিস নিয়ে একটা বস্তু তৈরি করলে (যেখানে $d = 2, 3, \ldots$), সর্বোচ্চ এনট্যাঙ্গলমেন্ট-ওলা স্টেট হবে $U_A \otimes U_B |\Phi^+\rangle$, যেখানে $U_A$, $U_B$ হলো $\mathbb{C}^d$-এর ওপর দুটো ইউনিটেরি অপরেটর, আর $|\Phi^+\rangle = \frac{1}{\sqrt{d}} \sum_{i=0}^{d-1} |ii\rangle$। এখানে $\{|i\rangle\}_{i=0}^{d-1}$ হলো $\mathbb{C}^d$-এর একটা অর্থনর্মাল বেসিস।

এখনও পর্যন্ত সর্বোচ্চ এনট্যাঙ্গলমেন্ট আছে এমন স্টেট খুঁজতে আমরা ইনফর্মেশন তত্ত্ব ব্যবহার করেছি। আমরা বলেছি যে এই সর্বোচ্চ এনট্যাঙ্গলমেন্ট নিয়ে আমরা কিছু কাজ করতে পারব যা অন্য কোনো স্টেট নিয়ে করলে অত ভালোভাবে করা যেত না (বা করাই যেত না)। এই ব্যাপারটা আমরা শুধু বলেছি। দেখিনি এখনও। একটু পরেই দেখব। কিন্তু এইসব আমরা দু-অংশের বস্তুর জন্যেই বলছি। কোনো বস্তুর যদি দুটোর বেশি অংশ থাকে, তার জন্যে সর্বোচ্চ এনট্যাঙ্গলমেন্ট সংজ্ঞায়িত করাতে কিছু ফ্যাসাদ আছে। আমরা সেদিকে বিশেষ যাব না। তবু এটুকু এখানে বলে রাখা যাক যে, আমরা যদি কোন স্টেট ব্যবহার করে সব থেকে ভালোভাবে কোনো কাজ করা যাবে, এই দিকে মন দিই, তাহলে তিন বা ততোধিক অংশের বস্তুর ক্ষেত্রে, কাজ অনুযায়ী সবচেয়ে কাজের স্টেটটা বদলে যায়। অনেক কাজের ক্ষেত্রেই দেখা যায় যে

$$\frac{1}{\sqrt{2}} \left( |0\rangle^{\otimes n} + |1\rangle^{\otimes n} \right) \tag{7}$$

স্টেটটা সবচেয়ে ভাল। (এখানে আমরা একটা $n$-অংশের বস্তুর কথা বলছি।) আবার অন্য কাজের ক্ষেত্রে দেখা যায় যে

$$\frac{1}{\sqrt{n}} \left( \left|01^{\otimes(n-1)}\right\rangle + \left|101^{\otimes(n-2)}\right\rangle + \ldots + \left|1^{\otimes(n-1)}0\right\rangle \right) \tag{8}$$

স্টেটটা সবচেয়ে বেশি উপযোগী। এই দুটো ছাড়াও আরো স্টেট আছে যারা কোনো-কোনো কাজে বেশি কাজ দেয়। (7)-এর স্টেটটাকে গ্রীনবার্গার-হর্ন-ৎজাইলিঙা (GHZ) বা GHZ-মারমিন বা ক্যাট স্টেট বলে ডাকা হয়ে থাকে [8]। (8)-এর স্টেটটাকে ৎজাইলিঙা-হর্ন-গ্রীনবার্গার বা W স্টেট বলে [9]।

## ১০। কোয়ান্টম টেলিপোর্টেশন

এনট্যাঙলমেন্টকে ব্যবহার করা যেতে পারে, এই উপলব্ধিকে ইন্ধন জুগিয়েছে যে সব কাজ, তার মধ্যে সন্দেহাতীতভাবে অন্যতম হলো কোয়ান্টম টেলিপোর্টেশন। কমিক্সের চরিত্ররা যে-ভাবে চোখের নিমেষে এক জায়গা থেকে আরেক জায়গায় চলে যায়, সেভাবেই একটা বস্তুর কোয়ান্টম স্টেটকে আমরা দূরবর্তী অন্য একটা বস্তুতে নিয়ে যেতে পারি - কোয়ান্টম টেলিপোর্টেশন করে।

চোখের নিমেষে মানে কিন্তু তাৎক্ষণিক নয়! এ ব্যাপারে নিচে আরো কথা হবে।

এখানে এটা জানা-বোঝা একান্ত প্রয়োজন যে এই কাজ করতে কী কী কাঁচামাল লাগবে। এই ফর্দ আমরা বানাব, কিন্তু একটু পরে। প্রথমে ব্যাপারটা কী, সেটা জেনে নেওয়া যাক।

আকবরের কাছে একটা দু-ডামেনশনের বস্তু আছে, যার কোয়ান্টম স্টেটটা সে বীরবলের কাছে থাকা একটা বস্তুর স্টেট করে দিতে চায়। এই বস্তুদুটোর নাম দেওয়া যাক $A'$ আর $B$. আকবরের কাছে আরো একটা বস্তু আছে, যার নাম ধরা যাক $A$. আমরা যে কাজটা করতে চাই তার জন্যে $A$ আর $B$-কে (আলাদা আলাদা ভাবে) দু-ডামেনশনের বস্তু হলেই চলবে। কিন্তু তাদের $\mathbb{C}^2 \otimes \mathbb{C}^2$-র সর্বোচ্চ এনট্যাঙলমেন্ট-ওলা স্টেট হতে হবে, যেমন সিঙ্গলেট স্টেট। যদি সিঙ্গলেটই ধরে নিই, আর $A'$-এর স্টেটটা যদি $|\psi\rangle = a|0\rangle + b|1\rangle$ হয়, যেখানে $|0\rangle$ আর $|1\rangle$ দুটো অর্থনর্মাল স্টেট এবং $|\psi\rangle$ স্টেটটা নর্মালাইজ্ড, তাহলে দুজনের কাছে থাকা এই তিনটে বস্তুর গোটা স্টেটটা হলো

$$|\psi\rangle_{A'} \otimes |\psi^-\rangle_{AB}.$$

সহজেই দেখা যায় যে এই স্টেটটাকে আমরা এইভাবে লিখতে পারি:

$$\frac{1}{2} \sum_{i=1}^{4} |B_i\rangle_{A'B} \otimes \sigma_i |\psi\rangle_B,$$

যেখানে $|B_i\rangle$-গুলো $|\psi^\pm\rangle, |\phi^\pm\rangle$-এর মধ্যে থেকে নেওয়া (যাদের অনেক সময় বেল স্টেট বলা হয়), আর $\sigma_i$-গুলো আইডেন্টিটি আর পাওলি ম্যাট্রিক্সদের থেকে নেওয়া। অবশ্যই একটা বিশেষ ক্রমে বেল স্টেট আর ম্যাট্রিক্স গুলোকে নিতে হবে। ব্যস এইটুকু বীজগণিত করতে হয় কোয়ান্টম টেলিপোর্টেশন বুঝতে গেলে।

এখানে $|\psi^\pm\rangle = (|01\rangle \pm |10\rangle)/\sqrt{2}$ আর $|\phi^\pm\rangle = (|00\rangle \pm |11\rangle)/\sqrt{2}$.

এইবার আকবরকে তার ল্যাবে থাকা দুটো অংশের ওপর, মানে $A'$ আর $A$-র ওপর, একটা পরিমাপ করতে হবে। বেল বেসিসে, অর্থাৎ $\{|B_i\rangle\}$ বেসিসে। ধরা যাক যে পরিমাপ

করে আকবর $|B_k\rangle$ পেল, যেখানে $k = 1$ বা $2$ বা $3$ বা $4$. এই সংখ্যাটাকে তাকে ফোন করে বীরবলকে বলতে হবে। ফোন পাওয়ার পর বীরবলের কাছে যে স্টেটটা আছে, সেটা হলো

$$\sigma_k|\psi\rangle.$$

বীরবল জানে যে পাওলি ম্যাট্রিক্সগুলো ইউনিটেরি আর $\sigma_k^2 = \mathbb{I}_2 \; \forall k$. তাই সে যদি তার অংশের ওপর $\sigma_k$ চালিয়ে দেয়, তাহলে তার স্টেটটা হয়ে যাবে $|\psi\rangle$, যেটা কিনা আকবরের ল্যাবে থাকা $A'$ অংশের স্টেট ছিল, আকবর $A'A$-এর ওপর বেল পরিমাপ (অর্থাৎ বেল বেসিসে পরিমাপ) চালানোর আগে। একেই বলে কোয়ান্টম টেলিপোর্টেশন। ১৯৯৩ সালে এটা আবিষ্কৃত হয়, তাত্ত্বিক দিক থেকে [10]। সত্যিকারের বস্তুত্রয়ের মধ্যে এ জিনিস করতে আরো বছর চার-পাঁচ সময় লাগে।

কোয়ান্টম টেলিপোর্টেশনকে তাত্ত্বিকভাবে বুঝতে, বীজগণিত একেবারেই ছোটখাটো সাদাসিধে হলেও, লক্ষ্য করার বিষয় অনেক। তাদের কয়েকটা নিচে আলোচনা করা হলো।

### কোয়ান্টম টেলিপোর্টেশন বনাম বিশেষ আপেক্ষিকতা

কোয়ান্টম টেলিপোর্টেশন অবশ্যই তাৎক্ষণিক নয়! আকবরের বেল পরিমাপ আর বীরবলের পাওলি চালানোর জন্যে যদি কোনো সময় নাও লাগে, আকবর বীরবলকে তার বেল পরিমাপের ফলাফল জানানোর জন্যে ফোন বা ওইরকম কিছু ব্যবহার করতে হবে, যেটা বড়জোর আলোর গতিতে যাবে। এখানে আমরা বিশেষ আপেক্ষিকতা তত্ত্বকে সত্যি বলে ধরে নিয়েছি, তবে শুধুমাত্র দুজনের ভেতরের ফোনালাপের ক্ষেত্রে। স্থানীয় কাজগুলো, যথা বেল পরিমাপ আর পাওলি চালানো, শুধু তাত্ত্বিকভাবে ভাবলে, ওরা যত খুশি তাড়াতাড়ি করতে পারবে - যত তাড়াতাড়ি করতে চাইবে, ততো বেশি শক্তির প্রয়োজন পড়বে। বস্তুর শক্তি বেশি বেড়ে গেলে আমাদের কোয়ান্টম ফীল্ড ব্যবহার করতে হবে, যেদিকে আমরা এখানে যাচ্ছি না।

উল্টোদিকে, যদি আমরা ধরে নিই যে কোনো একটা উপায়ে আকবরের কাছ থেকে বীরবলের কাছে একটা কোয়ান্টম স্টেট চলে যেতে পারে, কোনো কোয়ান্টম বা ক্লাসিকল চ্যানেল ব্যবহার না করে [11], তাহলে আমরা বিশেষ আপেক্ষিকতা তত্ত্বকে লঙ্ঘন করতে পারব। এটা দেখানোর জন্যে, ধরা যাক আকবর আর বীরবল আগে থেকে ঠিক করেছে যে আকবরের ল্যাবের কাছে আগামী ১৫ই আগস্ট দুপুর বারোটায় আকাশে মেঘ থাকলে সে বীরবলকে $|0\rangle$ পাঠাবে, আর অন্যথায় $|1\rangle$ পাঠাবে। তাছাড়া, ধরা যাক যে ওরা দুজনে এতটাই দূরে আছে যে আলোকে এক জায়গা থেকে অন্যটাতে যেতে পাঁচ সেকন্ড সময় লাগে। এবার ওরা এমন একটা কোয়ান্টম প্রোটকল ব্যবহার করল, যাতে ওদের দুজনের ভেতরের কোনো কোয়ান্টম বা ক্লাসিকল চ্যানেলের দরকার পড়ছে না, এবং যাতে তাদের সমস্ত স্থানীয় কাজকর্ম দু সেকন্ডে শেষ হয়ে যায়। তবে, আকবর-বীরবলের কাছে আগে থেকেই কোনো দু-অংশের বস্তু থাকতে পারে, যেটা এনট্যাঙ্গল্ড এবং যেটার একটা অংশ আকবরের কাছে আর অন্যটা বীরবলের কাছে। এখন $|0\rangle$ আর $|1\rangle$-দের বীরবল চিনতে পারবে, কারণ ওরা অর্থাগনল। তাহলে বারোটা বেজে দু সেকন্ডেই বীরবল জেনে যাবে যে আকবরের ল্যাবের কাছের আকাশ মেঘলা, না তা নয়। এটা অবশ্যই বিশেষ আপেক্ষিকতার সুস্পষ্ট লঙ্ঘন।

এদিকে ওপরে বর্ণনা করা কোয়ান্টম টেলিপোর্টেশন প্রোটকালটা ব্যবহার করা হলে কিন্তু ফোনালাপটা বাদ দিয়ে দিলে, বীরবল জানতে পারবে না, আকবরের বেল পরিমাপের কী ফল হয়েছিল, আর তাই বীরবলের কাছে থাকা অংশটার স্টেট, বীরবলের কাছে, হয়ে থাকবে $\frac{1}{2}\mathbb{I}_2$, যা কিনা বেল পরিমাপ করার আগেই ছিল। কারণ, বেল পরিমাপের আগে, বীরবলের কাছে থাকা অংশটার স্টেট ছিল $\text{tr}_A|\psi^-\rangle\langle\psi^-|$, যেটা হলো $\frac{1}{2}\mathbb{I}_2$. আর বেল পরিমাপের পরে, আকবরের কাছ থেকে কোনো সংবাদ পাওয়ার আগে পর্যন্ত, বীরবলের কাছে থাকা অংশটার

স্টেট, বীরবলের মতে

$$\frac{1}{4} \sum_{i=1}^{4} \sigma_i |\psi\rangle\langle\psi| \sigma_i,$$

যেটা আবার সেই $\frac{1}{2}\mathbb{I}_2$, সে $|\psi\rangle$ যাই হোক না কেন।

### কাঁচামালের ফর্দ

কোয়ান্টম টেলিপোর্টেশন মারফত আমরা একটা বস্তুর কোয়ান্টম স্টেটকে এক জায়গা থেকে আরেক জায়গায় পাঠিয়ে দিতে পারি। তা এইবেলা দেখে নেওয়া ভালো এটা করতে কী কী জিনিস জরুরি। ফোকটে আর কী পাওয়া যায়!?

প্রথমে আমরা একটা দু-ডামেনশনের বস্তুর কোয়ান্টম স্টেটের টেলিপোর্টেশনের কথাই বলি। দেখানো যায় যে কোয়ান্টম তত্ত্ব মেনে এটা করতে, অন্তত একটা সিঙ্গলেট স্টেট (বা তার সাথে স্থানীয় ইউনিটেরি দিয়ে সম্পর্কিত একটা স্টেট) লাগবেই [12]। অবশ্যই এই একই কাজ অন্য কোনো দু-অংশের বস্তুর স্টেট ব্যবহার করেও করা যায়, যাকে LOCC দিয়ে, ডিটরমিনিস্টিকালি (অর্থাৎ, নিশ্চিতভাবে, প্রাবিলিটি 1 সহ), সিঙ্গলেট স্টেটে নিয়ে যাওয়া চলে, যেমন

$$\frac{1}{\sqrt{3}} \left(|00\rangle + |11\rangle + |22\rangle\right), \tag{9}$$

বা

$$\frac{1}{\sqrt{2}}|00\rangle + \frac{1}{\sqrt{3}}|11\rangle + \frac{1}{\sqrt{6}}|22\rangle, \tag{10}$$

বা

$$p|\phi^+\rangle\langle\phi^+| + (1-p)|\tilde{\phi}^+\rangle\langle\tilde{\phi}^+|, \tag{11}$$

যেখানে $0 \leq p \leq 1$ আর $|\tilde{\phi}^+\rangle = (|02\rangle + |13\rangle)/\sqrt{2}$. এখানে $\{|0\rangle, |1\rangle, |2\rangle\}$ এবং $\{|0\rangle, |1\rangle, |2\rangle, |3\rangle\}$-রা মিউচুয়লি অর্থনর্মাল স্টেটের সেট। (9) আর (10)-এর স্টেটগুলো থেকে যে ডিটরমিনিস্টিক LOCC দিয়ে সিঙ্গলেটে পৌঁছনো যায়, সেটা বুঝতে একটু কাঠখড় পোড়াতে হবে, যা আমরা এখানে করব না। শুধু বলে রাখা যাক যে এটা যে হয়, তা জানতে এই পেপরগুলোর যেকোন একটা পড়া যায়: [13]। (11)-এর স্টেটটাকে যে সিঙ্গলেট স্টেটে নিয়ে যাওয়া যায়, ডিটরমিনিস্টিক LOCC দিয়ে, সেটা বোঝা অপেক্ষাকৃত অনেক সহজ। আসলে (11)-এর স্টেটটা (যেকোন $p$-এর জন্যে) "লুকিয়ে লুকিয়ে" একটা পিওর স্টেট [14]।

আর ওপরে আমরা বিশদে আলোচনা করেছি যে একটা ক্লাসিকল চ্যানল লাগবে। আমরা দেখেছি যে দুটো ক্লাসিকল বিট পাঠানো যথেষ্ট। কিন্তু তার কমে (কিন্তু শূন্যের চেয়ে বেশি দিয়ে) কি কিছু করা যায়? উত্তর হলো "না", কিন্তু এই উত্তরের প্রমাণ বুঝতে আমাদের কোয়ান্টম ডেন্স কোডিং [15] জানতে হবে। আমরা এখানে সেদিকে যাব না।

### কোয়ান্টম ইনফর্মেশন ক্লোন করা যায় না ও তার সাথে কোয়ান্টাম টেলিপোর্টেশনের সম্পর্ক

কোয়ান্টাম টেলিপোর্টেশন করে আকবরের কাছে থাকা $A'$-এর স্টেটটা তো বীরবলের জিম্মায় থাকা $B$-র স্টেট হয়ে গেল। কিন্তু $A'$-এর টেলিপোর্টেশন-উত্তর স্টেটটা কী দাঁড়াল দেখা যাক। টেলিপোর্টেশন হয়ে যাওয়ার পর, আকবরের গবেষণাগারের $A'$ আর $A$ অংশদুটো কোনো একটা বেল স্টেটে থাকবে, আর তাই $A'$-এর স্টেটটা থাকবে $\frac{1}{2}\mathbb{I}_2$ স্টেটে। তাহলে আমরা

দেখছি যে আকবরের ল্যাবে $|\psi\rangle$ স্টেটটার, অর্থাৎ $a, b$-র চিহ্নমাত্র নেই। আকবর আর বীরবলের ল্যাবগুলোতে এবং ওদুটোর মধ্যের ক্লাসিকল চ্যানেলটাতে ছাড়া, বিশ্ব-চরাচরের আর কোথাও আমরা টেলিপোর্টেশন করতে গিয়ে কিছু করিনি, এবং তাই আর কোথাও $a, b$-র ছাপ থাকবে না। ক্লাসিকল চ্যানেলটাতে অত জায়গা নেই। মাত্র দুটো বিট ধরে, এমন ক্লাসিকল চ্যানেল ব্যবহার করা হয়েছে। এদিকে $a$ আর $b$-র ভেতর দুটো বাস্তব সংখ্যা আছে, যাদের ক্লাসিকল বিট দিয়ে প্রকাশ করতে গেলে অসংখ্য বিটের প্রয়োজন। সুতরাং কোয়ান্টম টেলিপোর্টেশন সমাধা হওয়ার পর, শুধুমাত্র বীরবলের ল্যাবে এক পিস $|\psi\rangle$ আছে। জগতের আর কোথাও ও জিনিস নেই।

$|\psi\rangle$ এবং $|\psi\rangle = a|0\rangle + b|1\rangle$, এবং $a, b$-কে ক্লাসিকল বিট মারফত প্রকাশ করতে চাইলে কী হয়, সেই নিয়ে কিছু কথা এইবেলা বলে নেওয়া যাক। কোয়ান্টম তত্ত্বের লিনিয়ারিটি দাবি করে যে $|\psi\rangle$ যদি একটা দু-ডামেনশনের বস্তুর স্টেট হয়, এবং সেই স্টেটের হিলবার্ট স্পেসের একটা অর্থনর্মাল বেসিস যদি $\{|0\rangle, |1\rangle\}$ হয়, তাহলে $a$ এবং $b$ যেকোন জটিল সংখ্যা হতে পারে। বর্ন নিয়ম মেনে চলার জন্যে আমাদের শুধু সেই সব $a, b$ নেওয়াই যথেষ্ট, যারা $|a|^2 + |b|^2 = 1$ মেনে চলে। তাছাড়া, সামগ্রিক একটা ফেজ আমরা বাদ দিতে পারি। এই সব মেনে টেনে নেওয়ার পর আমরা $a$ আর $b$-কে যথাক্রমে $\cos(\theta/2)$ এবং $e^{i\phi} \sin(\theta/2)$ দিয়ে বদলে দিতে পারি, যেখানে $\theta \in [0, \pi]$ আর $\phi \in [0, 2\pi)$।

যে বস্তুকে বর্ণনা করার জন্যে একটা দু-ডামেনশনের হিলবার্ট স্পেসের প্রয়োজন হয়, তাকে কোয়ান্টম ইনফর্মেশন রিসর্চরদের পরিভাষায় "**কিউবিট**" বলা হয়ে থাকে, যা কিনা "কোয়ান্টম বিট" থেকে সংক্ষেপিত। কোয়ান্টম ফিকির আছে এমন যেকোন যন্ত্রের মৌলিক একক হলো একটা কিউবিট। ঠিক যেমন ক্লাসিকল যেকোন যন্ত্রের মৌলিক একক হলো একটা বিট। তাহলে আমরা ওপরে যা দেখলুম তা হলো, একটা কিউবিটের স্টেটকে আমরা একটা গোলক দিয়ে প্রকাশ করতে পারি, যেমন একটা ফুটবলের বাইরের যে অংশটা দেখা যায়, সেটা দিয়ে। এই গোলকের ব্যাসার্ধ $1$ হতে হবে। $\theta$ এবং $\phi$ রেডিয়নে মাপা হবে, কিন্তু ব্যাসার্ধের কোনো একক থাকবে না। একটা নিরেট ফুটবল যদি নেওয়া হয়, তাহলে দু-ডামেনশনের বস্তুর মিক্সড স্টেটগুলোকে ওই ফুটবলের ভেতরের অংশে পাওয়া যাবে। এই গোলকটাকে ব্লহ্‌ গোলক নামে ডাকা হয়। নিরেট গোলকটাকেও অনেক সময় একই নামে ডাকা হয়, বা ব্লহ্‌ বল বলা হয়। পঁয়কারে গোলক বলে একটা জিনিস আছে, যেটা এই ব্লহ্‌ গোলকের খুবই কাছাকাছি জিনিস, কিন্তু আমরা সেটা নিয়ে আর আলোচনা করব না।

আমরা আবার কোয়ান্টম টেলিপোর্টেশনে ফিরে যাই। আমরা ইতিপূর্বে দেখেছি যে কোয়ান্টম টেলিপোর্টেশন শুরু হওয়ার আগে যে স্টেটটা আকবরের কাছে ছিল, সেটা টেলিপোর্টেশনের পরে তার ল্যাব থেকে হাওয়া হয়ে গেছে, এবং বীরবলের ল্যাবে উদয় হয়েছে, এবং ভূ-ভারতে আর কোথাও আবির্ভূত হয় নি। তা, এ তো এই বিশেষ টেলিপোর্টেশন প্রোটকালে। অন্য কোনো প্রোটকালে কি দুই বা তার বেশি জায়গায় আবির্ভাব হতে পারত? এর উত্তর "না"। কারণটা নিহিত আছে একটা কোয়ান্টম "নো-গো" ফলাফলে। নিচে সেই নিয়ে ছোট করে আলোচনা করা হলো।

ধরা যাক আমরা একটা ছবির ফটোকপি তৈরি করব। তা গেলুম সেই জন্যে একটা ফটোকপি করার দোকানে। দোকানদারমশাই ওনার মেশিনে আমাদের আসল ছবিটা দিলেন, এবং সেই সঙ্গে দিলেন একটা খালি কাগজ। এবং মেশিন চালিয়ে দিলেন। এবং তারপর আমাদের দুটো কাগজ ধরিয়ে দিলেন - আসলটা এবং তার থেকে করা নকলটা। গোটা বিষয়টা যদি আমরা কোয়ান্টম তত্ত্বের আলোয় দেখতে চাই, তাহলে আমাদের আসল ছবিটার একটা কোয়ান্টম স্টেট নিতে হবে। ধরা যাক সেটা $|\psi\rangle$. কোনো একটা ফটোকপি করার যন্ত্র তখনই কাজের, যখন তা দিয়ে যেকোন ছবি নকল করা যায়। তাই আমরা ধরে নেব যে আসলটার একটা স্টেট থাকলেও, আমরা সেটা জানি না। এদিকে খালি কাগজটা আমরাই (মানে, দোকানী) ফটোকপির যন্ত্রে দিচ্ছি, আর এই একই কাগজ প্রত্যেকবার দেব - যেকোন ছবি কপি করার জন্যে। তাই আমরা ধরে নেব যে এটার স্টেট আমরা জানি। ধরা যাক সেটা $|B\rangle$. আর যন্ত্রটা ধরা যাক নকল করার আগে $|M\rangle$ স্টেটে আছে। তাহলে একটা তিন-অংশের বস্তুর কথা

আমরা ভাবছি, যেটা আছে এই স্টেটে:

$$|\psi\rangle \otimes |B\rangle \otimes |M\rangle.$$

মেশিন থেকে যা বেরবে, তা আমরা চাই এই স্টেটে:

$$|\psi\rangle \otimes |\psi\rangle \otimes |M_\psi\rangle.$$

অর্থাৎ আসলটা ফেরত চাই, নকলটা একেবারে আসলের মতো চাই, আর যন্ত্রটা এই ফাঁকে একটু বদলে গিয়ে থাকতে পারে, এবং এই বদলটা আসল স্টেটটার ওপর নির্ভর করতে পারে। এদিকে আসলের স্টেটটা তো জানি না, আর তাই সেটা $|\psi\rangle$ না হয়ে $|\phi\rangle$-ও হতে পারে। তাই আমরা চাই যে একই যন্ত্রে নিচের দুটো পরিবর্তনই হোক:

$$|\psi\rangle \otimes |B\rangle \otimes |M\rangle \quad \to \quad |\psi\rangle \otimes |\psi\rangle \otimes |M_\psi\rangle,$$
$$|\phi\rangle \otimes |B\rangle \otimes |M\rangle \quad \to \quad |\phi\rangle \otimes |\phi\rangle \otimes |M_\phi\rangle.$$

কোয়ান্টম তত্ত্ব বলে যে যন্ত্রমাত্রই ইউনিটেরি মতে চলে, এবং যেকোন ইউনিটেরি পরিবর্তনে ইনপুটদের এবং আউটপুটদের ইনার প্রোডাক্ট অপরিবর্তিত থাকে। সুতরাং আমাদের কোয়ান্টম copyকলে $|\psi\rangle$ এবং $|\phi\rangle$ কপি করা গেলে, এই সম্পর্কটা থাকতেই হবে:

$$\langle \psi | \phi \rangle = \langle \psi | \phi \rangle^2 \langle M_\psi | M_\phi \rangle. \tag{12}$$

কোশি-শ্ভাৎস অসমীকরণ ব্যবহার করে দেখানো যায় যে সমীকরণ (12)-র বৈধ হওয়ার দুটো উপায় আছে। এক, যদি $|\psi\rangle$ আর $|\phi\rangle$ সমান হয়, অর্থাৎ আমাদের যন্ত্র মাত্র একটাই স্টেট কপি করতে পারবে - এই যন্ত্র আমরা আগেই ফালতু বলে নাকচ করে দিয়েছি। আর দুই, যদি $|\psi\rangle$ এবং $|\phi\rangle$ অর্থাগনল হয়। সুতরাং কোয়ান্টম তত্ত্ব যদি সত্যি হয়, তাহলে আমরা একটা কল ব্যবহার করে শুধুমাত্র একটা মিউচুয়ালি অর্থাগনল স্টেটের সেটকেই কপি করতে পারব। দুটো নন-অর্থাগনল স্টেটকে কপি করতে পারে এমন যন্ত্র কোয়ান্টম তত্ত্ব মেনে চলে তৈরি করা যায় না। একেই বলে কোয়ান্টম নো-ক্লোনিং রেজাল্ট, এবং অনেকগুলো কোয়ান্টম নো-গো রেজাল্টের একটা [16]।

এতদিন যে সব দোকানিরা আমাদের কাছ থেকে ফটোকপি করার পর পয়সা নিতেন, তাঁরা কি তাহলে ঠকিয়ে নিতেন? ঠকে এবং ঠেকেই এ জগতে সব কিছু শিখতে হয়, কিন্তু এখানে ব্যাপারটা অন্যরকম। দোকানদারমশাইদের কাছে থাকা যন্ত্রগুলো যেসব কাগজের নকল করতে পারে, তারা সবাই একে অন্যের সাথে অর্থাগনল।

তাহলে আমরা দেখছি যে কোয়ান্টম টেলিপোর্টেশনের শেষে একটা মাত্র $|\psi\rangle$-এর কপি পড়ে থাকাটা কোনো আকস্মিক ঘটনা না - কোয়ান্টম নো-ক্লোনিং উপপাদ্য দ্বারা এরকমটাকে বাধ্য করা হয়েছে, এর অন্যথা হতে পারত না। তবে এটা তখনই সত্যি যখন আমরা $|\psi\rangle$ অর্থাৎ $a, b$ জানি না। চেনা স্টেটকে তো ক্লোন করা যায়। সুতরাং চেনা স্টেটের ক্ষেত্রে আকবর $|\psi\rangle$ স্টেটটার অনেককটা কপি তৈরি করতে পারবে, এবং যেকোন একটা কপি বীরবলকে পাঠিয়ে দিতে পারবে - কোয়ান্টম টেলিপোর্ট করে। একটার বেশি পাঠাতে গেলে কিন্তু আরো কাঁচামাল লাগবে, অর্থাৎ আরো সিঙ্গলেট এবং আরো ক্লাসিকল বিট লাগবে।

কিন্তু চেনা স্টেট পাঠানোর জন্যে সিঙ্গলেট ব্যবহার করার দরকারটা কী? শুধু ক্লাসিকল বিট ব্যবহার করা যায় না? যায়, কিন্তু অসংখ্য বিটের দরকার হবে। আনকাউন্টেব্লি ইনফিনিট বিটের প্রয়োজন হবে। তার কারণ $|\psi\rangle$-এর খবর বীরবলকে বলতে গেলে $\theta$ আর $\phi$ পাঠাতে হবে, যারা দুটো বাস্তব সংখ্যা, যাদের যেকোন একটাকে ক্লাসিকল বিট প্রকাশ করতেই আনকাউন্টেব্লি ইনফিনিট বিটের প্রয়োজন হয়।

আচ্ছা, যদি এমন হতো যে আমরা এমন একটা দুনিয়াতে থাকি, যেখানে ক্লোনিং করা যায়। তাহলে আকবর তার $A'$-এর স্টেটটাকে অনেকবার ক্লোন করে ফেলতে পারত, এবং

তারপর তাদের ওপর বিভিন্ন পরিমাপ করে $|\psi\rangle$-কে, অর্থাৎ $a$, $b$-কে, অর্থাৎ $\theta$, $\phi$-কে চিনে ফেলতে পারত। [একই কোয়ান্টম স্টেটের অনেক কপি থেকে স্টেটটাকে চিনে ফেলার পদ্ধতিকে কোয়ান্টম টমোগ্রফি বলে, যেটা নিয়ে আমরা আলোচনা করব না।] এবং তারপর ক্লাসিকল চ্যানল ব্যবহার করে বীরবলকে $\theta$, $\phi$ কী, তা বলে দিতে পারত। সুতরাং এক্ষেত্রে টেলিপোর্ট করার জন্যে সিঙ্গলেট স্টেটটার আর প্রয়োজন হতো না। তাহলে আমরা দেখছি: যে দুনিয়ায় নো-ক্লোনিং বলবৎ নয়, সেখানে টেলিপোর্টেশন করতে শুধু ক্লাসিকল চ্যানলই যথেষ্ট। অর্থাৎ কাঁচামালের ফর্দতে শেয়ার হয়ে থাকা এনট্যাঙ্গল্ড স্টেটের বদলে একটা নো-ক্লোনিং জারি নেই এরকম জগৎ বসিয়ে দেওয়া চলবে [17]। কিন্তু তখন সেই ফর্দের দু নম্বর আইটেমে দুটো ক্লাসিকল বিটের বদলে আনকাউন্টেব্লি ইনফিনিট বিট রাখতে হবে।

## ১১। থেকে ১৮। পরের ভার্শনে থাকবে।

## ১৯। শেষের কথা

এনট্যাঙ্গলমেন্ট কোয়ান্টম স্টেটের একটা মৌলিক ধর্ম। প্রতিদিনই নতুন নতুন ভৌত অবস্থা ও ঘটনা বুঝতে আর জানতে এনট্যাঙ্গলমেন্টকে কাজে লাগানো হচ্ছে। মাঝে মাঝেই নতুন উপযোগ খুঁজে পাওয়া যাচ্ছে। এইটুকু একটা আলোচনায় সেই সব কিছু তো বলা যায় নি বটেই, কোনো একটা সমীক্ষাতে গোটা ব্যাপারটা আদৌ ধরা যেতে পারে কিনা সে নিয়ে সন্দেহ আছে। তবে এই আলোচনার তুলনায় আরো অনেক দীর্ঘ ও ব্যাপ্ত সমীক্ষা বাজারে আছে ... সেগুলোর ফিরিস্তি আর এখানে দেওয়া হলো না ... তাদের প্রায় সবকটাই `arXiv.org`-এ আছে।

এনট্যাঙ্গলমেন্টের বয়েস প্রায় নব্বুই ছুঁইছুঁই। কিন্তু এনট্যাঙ্গলমেন্টকে কাজে লাগানো যাবে, এই উপলব্ধি বছর তিরিশেকের। সুতরাং কোয়ান্টম প্রযুক্তি সতিযই মানুষের কাজে লাগবে কিনা, সেটা বোঝার সময় যেমন এখনও শেষ হয় নি, তেমন দেরিও এখনও হয় নি।

## কৈফিয়ত

এই লেখা পড়ে কেউ বাংলায় রিসর্চ পেপর লিখতে চাইবে, এরকমটা মোটেই আশা করা হয় নি এখানে। এনট্যাঙ্গলমেন্টের একেবারে গোড়ার ব্যাপারটা ইংরেজিতে পড়ে বুঝতে আমার প্রথম-প্রথম বেশ অসুবিধে হয়েছিল। সেরকম আরো কিছুজন ভূভারতে বর্তমান, এই সন্দেহবশত এই ছোট লেখার খেয়াল। এবং আশা যে আমাদের উপকার করার জন্যে আরো কয়েকজন এই রকম কলম ধরবেন।

## তথ্যসূত্রসমূহ